\documentclass[11pt,a4paper]{article}
\pdfoutput=1

\usepackage[utf8]{inputenc}
\usepackage[T1]{fontenc}
\usepackage[]{graphicx}
\graphicspath{{../Figures/}{./}}

\usepackage{subfig, caption, a4wide, fancyvrb, upquote}
\usepackage{longtable, multirow, enumitem}
\usepackage{amsmath, amsfonts, amssymb}
\usepackage{chicago}
\usepackage[margin=1in]{geometry}

\let\bibhang\relax
\usepackage[authoryear]{natbib}
\newcommand{\T}{{}^{\top}}
\newcommand{\x}{\boldsymbol{x}}

\newcommand{\y}{\boldsymbol{y}}
\newcommand{\z}{\boldsymbol{z}}
\newcommand{\J}{\boldsymbol{J}}

\newcommand{\A}{\boldsymbol{A}}
\newcommand{\Space}{\mathcal{S}}
\newcommand{\XSpace}{\mathcal{X}}
\newcommand{\YSpace}{\mathcal{Y}}
\DeclareMathOperator{\Real}{\mathbb{R}}
\newcommand{\thetab}{\boldsymbol{\theta}}

\newcommand{\mub}{\boldsymbol{\mu}}
\newcommand{\lambdab}{\boldsymbol{\lambda}}
\newcommand{\Lambdab}{\boldsymbol{\Lambda}}

\newcommand{\Sigmab}{\boldsymbol{\Sigma}}
\newcommand{\Psib}{\boldsymbol{\Psi}}

\renewcommand{\hat}[1]{\widehat{#1}}

\newcommand{\ISE}{\mathrm{ISE}}

\DeclareMathOperator{\Exp}{\mathrm{E}}
\DeclareMathOperator{\vech}{vech}
\DeclareMathOperator*{\argmax}{arg\max}

\makeatletter
\newcommand\code{\bgroup\@makeother\_\@makeother\~\@makeother\$\@codex}
\def\@codex#1{{\normalfont\ttfamily\hyphenchar\font=-1 #1}\egroup}
\makeatother

\begin{document}

\title{A transformation-based approach to Gaussian\\
       mixture density estimation for bounded data}
\author{%
Luca Scrucca\\
Universit\`a degli Studi di Perugia, Italy}
\date{\today}
\maketitle

\begin{abstract}
Finite mixture of Gaussian distributions provide a flexible semi-parametric methodology for density estimation when the variables under investigation have no boundaries. However, in practical applications variables may be partially bounded (e.g. taking non-negative values) or completely bounded (e.g. taking values in the unit interval). In this case the standard Gaussian finite mixture model assigns non-zero densities to any possible values, even to those outside the ranges where the variables are defined, hence resulting in severe bias. 
In this paper we propose a transformation-based approach for Gaussian mixture modelling in case of bounded variables. The basic idea is to carry out density estimation not on the original data but on appropriately transformed data. Then, the density for the original data can be obtained by a change of variables. Both the transformation parameters and the parameters of the Gaussian mixture are jointly estimated by the Expectation-Maximisation (EM) algorithm. The methodology for partially and completely bounded data is illustrated using both simulated data and real data applications. \\

\noindent{\it Keywords:} Bounded support; Density estimation; EM algorithm; Gaussian mixture models; Range-power transformation.
\end{abstract}

\newpage
\baselineskip=18pt

\section{Introduction}
\label{sec:intro}

Density estimation is the problem of inferring a probability density function given a finite number of sample data points drawn from a population described by a probability distribution.
Broadly speaking, three alternative approaches to density estimation can be distinguished. In the parametric approach a parametric distribution is assumed for the density with unknown parameters which are estimated by fitting the parametric function using the observed data.
Conversely, in the nonparametric approach no density function is assumed a priori, but its form is completely determined by the data. Histograms and kernel density estimation (KDE) are two popular methods that belong to this class, and both are characterised by the number of parameters growing with the size of the dataset. Furthermore, extensions to higher dimensionality is problematic.
A third approach is based on finite mixture models, where the unknown density is expressed as a convex combination of one or more probability density functions. In this class a popular model is the Gaussian mixture model which assumes the Gaussian distribution for the underlying component densities. Gaussian mixture models (GMMs) can approximate any continuous density with arbitrary accuracy provided the model has a sufficient number of components and the parameters of the model are correctly estimated \citep{Escobar:West:1995, Roeder:Wasserman:1997}.

\begin{figure}[htb]
\centering
\includegraphics[width=0.7\textwidth]{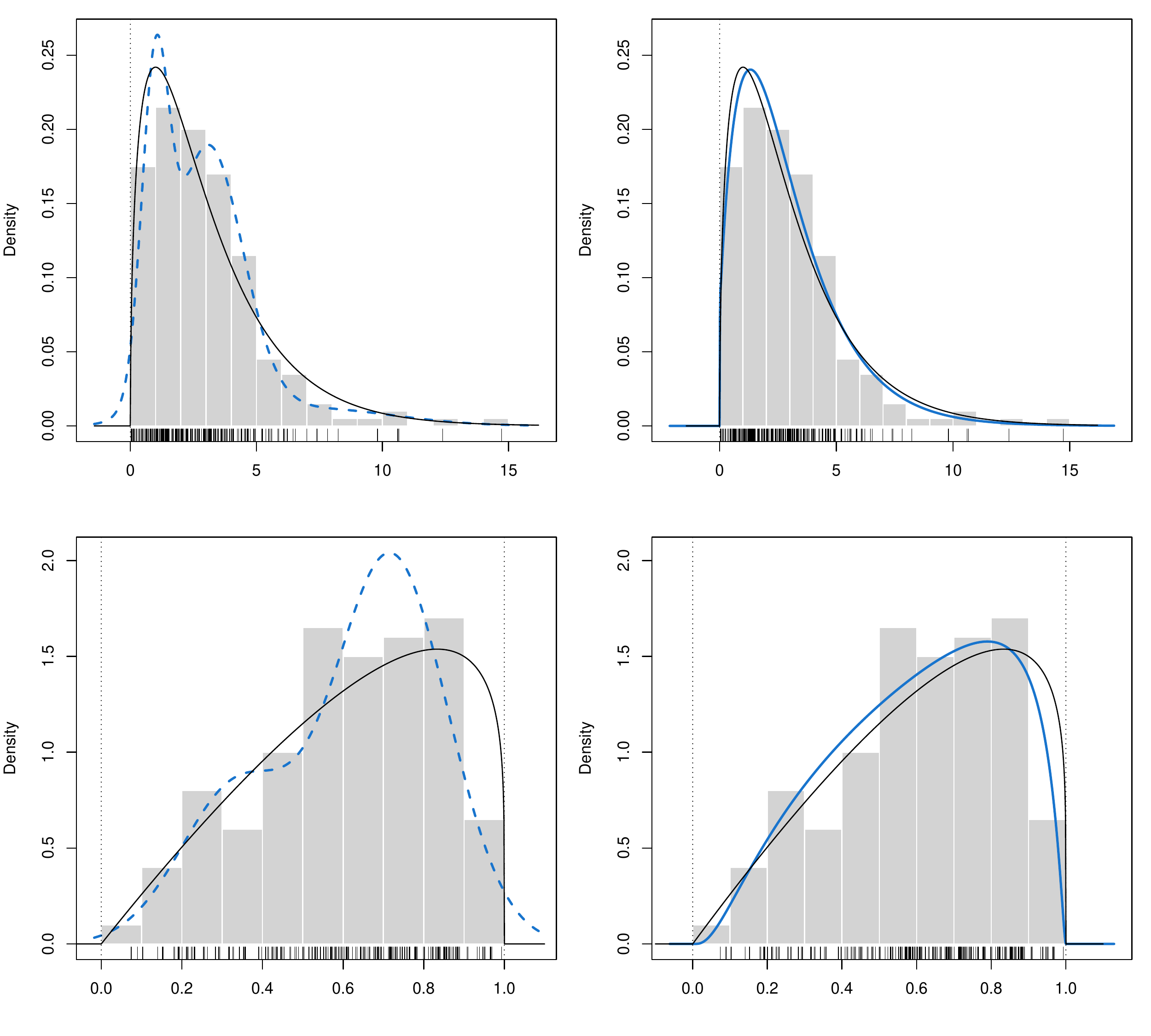}
\caption{\small Histograms for random samples drawn from a $\chi^2(3)$ (top panels) and a Beta$(2, 1.2)$ (bottom panels) distributions with the corresponding density functions (solid lines) and boundaries of the random variable (vertical dotted lines). Panels on the left show the estimated densities obtained by fitting a GMM on the original scale (blue dashed lines). Panels on the right show the densities estimated by the GMDEB transformation approach (blue thick lines).}
\label{fig:example1}
\end{figure}

Bounded data are quite common in biomedical data analyses because of the measurement scale of the data, or the type of variables under study.
However, the standard GMM for density estimation does not take into account whether or not a variable has bounded support. 
Consider the graphs in Figure~\ref{fig:example1} which show some histograms for random samples drawn from two distributions, one bounded from below (top panels), and one having both lower and upper bounds (bottom panels). In these graphs boundaries are shown as vertical dots, true densities are represented as solid lines, and density estimates based on GMMs as dashed lines (see left panels of Figure~\ref{fig:example1}). In both cases, the estimated densities are unsatisfactory at the boundaries, but also in the range of admissible values. A possible way to tackle these problems is to abandon the use of GMMs in favour of alternative component distributions. Another option is to remain in the realm of the Gaussian mixtures framework, but analyse the data in a transformed scale. The right panels of Figure~\ref{fig:example1} show the density estimates obtained with the GMDEB approach proposed in this paper. In both cases, the true underlying densities appear to be well approximated, and with natural boundaries constraints clearly satisfied.

In this paper a transformation-based approach to density estimation based on GMMs is proposed and discussed. 
The basic idea is to use an invertible function to map a bounded variable to an unbounded support, estimate the density of the transformed variable, and then back-transform to the original scale. 
This approach seems very natural and it has been around for a long time \citep[see][for KDE]{Wand:Marron:Ruppert:1991, Marron:Ruppert:1994}, but a simple and efficient implementation of this methodology is not yet available in the context of mixture density estimation.
Note that a similar approach based on Manly transformation has been recently proposed by \citet{Zhu:Melnykov:2018} for modelling skewed data in model-based clustering. However, our proposal differs in two main respects: firstly, it has been designed to allow variables with bounded support, secondly, it aims at a different goal, i.e. density estimation as compared to clustering. 

In Section 2 the GMMs approach to density estimation is reviewed. 
Then, Section 3 presents the proposed range-power transformation method for density estimation using GMMs in case of bounded variables. The model is described and the corresponding maximum likelihood estimates are derived through the EM algorithm. 
Section 4 contains the results of some simulation studies carried out to evaluate the proposed methodology and to compare with other available methods. In Section 5 some real-world datasets are analysed. The final section provides some concluding remarks.

\section{Finite mixture modelling}

\subsection{Finite mixture for density estimation}

Consider a vector of random variables $\x$ taking values in the sample space $\Space_\XSpace \subseteq \Real{^p}$ with $p \ge 1$, and assume that the probability density function can be written as a finite mixture density of $G$ components of the form
\begin{equation*}
f(\x; \Psib) = \sum_{g=1}^G \pi_g f_g(\x; \thetab_g),
\end{equation*}
where $\Psib = (\pi_1, \ldots, \pi_{G-1}, \thetab_1, \ldots, \thetab_G)\T$ is the parameters vector. The mixing weights $(\pi_1,\ldots,\pi_G)$ must satisfy the constraints $\pi_g > 0$ for all $g=1,\ldots,G$, and $\sum_{g=1}^G \pi_g = 1$. 
The $g$th component density $f_g(\x; \thetab_g)$ is usually taken as known except for the associated parameter(s) $\thetab_g$. Most applications assume that all component densities arise from the same parametric distribution family, although this need not be the case in general.
In particular, a popular model specifies $f_g(\x; \thetab_g) \equiv \phi(\x; \mub_g, \Sigmab_g)$, where $\phi(\cdot)$ is the Gaussian density with mean $\mub_g$ and covariance matrix $\Sigmab_g$. Then, the Gaussian mixture model (GMM) can be written as 
\begin{equation}
f(\x; \Psib) = \sum_{g=1}^G \pi_g \phi(\x; \mub_g, \Sigmab_g),
\label{eq:GMM}
\end{equation}
where in this case $\Psib = (\pi_1, \ldots, \pi_{G-1}, \mub_1\T, \ldots, \mub_G\T, \vech\{\Sigmab_1\}\T, \ldots, \vech\{\Sigmab_G\}\T)$ represents the entire parameters vector. Note that $\vech\{\cdot\}$ is an operator that forms a vector by extracting unique elements of a symmetric matrix.

In this paper we refer to \eqref{eq:GMM} as the Gaussian mixture density estimate (GMDE) model.
The usual nonparametric kernel density estimate (KDE) can be viewed as a mixture of $G = n$ components with uniform weights, i.e. $\pi_g = 1/n$ \citep[pp. 28--29]{Titterington:etal:1985}. Compared to KDE, finite mixture modelling uses a smaller number of components (i.e. less parameters), so it has smaller variance. Conversely, compared to parametric density estimation, finite mixture modelling has the advantage of (potentially) using more parameters, so introducing less estimation bias. There are also disadvantages related to mixture modelling, such as an increased learning complexity and lack of closed-form solution, so it needs to resort to numerical procedures (e.g. EM algorithm), and in certain cases there can be identifiability issues.

\subsection{Estimation of Gaussian finite mixture model}

Consider a random sample $\x_1, \ldots, \x_n$ of $n$ observations on $p$ variables drawn from the mixture distribution in \eqref{eq:GMM}. Then, the log-likelihood is given by
\begin{equation}
\ell(\Psib) = \sum_{i=1}^{n} \log
  \sum_{g=1}^G \pi_g \phi(\x_i; \mub_g, \Sigmab_g).
\label{eq:gmmloglik}
\end{equation}
Direct maximisation of the log-likelihood function is not straightforward, so MLEs are usually obtained via the Expectation-Maximisation (EM) algorithm \citep{Dempster:Laird:Rubin:1977, McLachlan:Peel:2000}.

An incomplete-data formulation of the mixture problem is introduced by associating to each observation a latent component-label vector $\z_i$ ($i=1,\ldots,n$). This is a $G$-dimensional vector, with the generic element $z_{ig} = 1$ or 0 according to whether or not $\x_i$ arises from the $g$th component of the mixture. 
Assuming independence of the complete-data vector $(\x_i\T,\z_i\T)\T$, and the multinomial distribution for the component-label vectors $\z_i$s, the complete-data log-likelihood is given by
\begin{equation}
\ell_C(\Psib) = \sum_{i=1}^{n} \sum_{g=1}^G z_{ig} 
                \{ \log\pi_g + \log\phi(\x_i; \mub_g, \Sigmab_g) \}.
\label{eq:gmmcloglik}
\end{equation}

The log-likelihood \eqref{eq:gmmloglik} is maximised using the EM algorithm, an iterative algorithm that alternates two steps, called E-step and M-step, which guarantees, under fairly general conditions, the convergence to at least a local maximiser. 
The objective function at iteration $(m+1)$ of the EM algorithm is the conditional expectation of the complete-data log-likelihood \eqref{eq:gmmcloglik}, the so-called Q-function:
\begin{equation*}
Q(\Psib;\Psib^{(m)}) = 
  \sum_{i=1}^{n} \sum_{g=1}^G \hat{z}_{ig}^{(m)}
                 \{ \log\pi_g + \log\phi(\x_i; \mub_g, \Sigmab_g) \},
\end{equation*}
where $\hat{z}_{ig}^{(m)} = \Exp(I(z_i=g)|\x_i, \Psib^{(m)})$, i.e. the estimated posterior probability at iteration $m$ of the EM algorithm, with $I(\cdot)$ the indicator function which equals 1 if the condition is fulfilled and 0 otherwise.

In the E-step the Q-function is evaluated, using the parameter values $\pi_g, \mub_g, \Sigmab_g$ obtained at the previous step, to get the updated posterior probabilities
\begin{equation*}
\hat{z}_{ig}^{(m+1)} = 
  \dfrac{\hat{\pi}_g^{(m)} \phi(\x_i; \hat{\mub}_g^{(m)}, \hat{\Sigmab}_g^{(m)})}
        {\sum_{k=1}^G \hat{\pi}_k^{(m)} \phi_k(\x_i; \hat{\mub}_k^{(m)}, \hat{\Sigmab}_k^{(m)})}.
\end{equation*}
Then, in the M-step the parameters vector $\Psib$ is updated by maximising the $Q$-function given the previous values $\hat{\Psib}^{(m)}$ and the updated posterior probabilities $\hat{z}_{ig}^{(m+1)}$, i.e.
\begin{equation*}
\hat{\Psib}^{(m+1)} = \argmax_{\Psib} Q(\Psib;\hat{\Psib}^{(m)}).
\end{equation*}
In the case of a multivariate Gaussian mixture the M-step yields
\begin{equation*}
\hat{\pi}_g^{(m+1)} = \dfrac{\sum_{i=1}^n \hat{z}_{ig}^{(m+1)}}{n}
\qquad\text{and}\qquad
\hat{\mub}_g^{(m+1)} = \dfrac{\sum_{i=1}^n \hat{z}_{ig}^{(m+1)} \x_i}{\sum_{i=1}^n \hat{z}_{ig}^{(m+1)}}.
\end{equation*}
The update formula for the covariance matrix depends upon the structure of the within-component covariance matrices. Parsimonious parameterisation of the covariance matrices can be expressed through the eigendecomposition $\Sigmab_g = v_g\O_g\A_g\O\T_g$, where $v_g$ is a scalar controlling the volume of the corresponding ellipsoid, $\A_g$ is a diagonal matrix specifying the shape of the density contours, and $\O_g$ is an orthogonal matrix which determines the orientation of the ellipsoid \citep{Banfield:Raftery:1993, Celeux:Govaert:1995}. For instance, assuming an unconstrained covariance matrix, the updating formula is
\begin{equation*}
\hat{\Sigmab}_g^{(m+1)} = 
  \dfrac{\sum_{i=1}^n \hat{z}_{ig}^{(m+1)}
         \left(\x_i - \hat{\mub}_{g}^{(m+1)}\right)
         \left(\x_i - \hat{\mub}_{g}^{(m+1)}\right)\T}
        {\sum_{i=1}^n \hat{z}_{ig}^{(m+1)}}.
\end{equation*}
\citet[][Table~3]{Scrucca:etal:2016} summarise some parameterisations of within-component covariance matrices, and the corresponding geometric characteristics, currently available in the \code{mclust} software. The previous unconstrained covariance matrix is indicated as VVV model. Note, however, that for some models no closed-formula is available, so numerical optimisation is required. 

The EM algorithm requires the specification of initial values for the the parameters, say $\Psib^{(0)}$. Alternatively, an initial assignment of observations to the components of the mixture can be made, basically starting the EM algorithm from the M-step. In any case, the initialisation of the EM algorithm is often crucial because the likelihood surface tends to have multiple modes, although it usually produces sensible results when started from reasonable starting values \citep[p. 150]{Wu:1983}. For a further discussion on this point and a recent proposal see \citet{Scrucca:Raftery:2015}.

Information criteria based on penalised forms of the log-likelihood are routinely used in finite mixture modelling for model selection, i.e. to decide how many components should be included in the mixture, but also which covariance parameterisations to adopt in the Gaussian case.
Two popular criteria are the Bayesian information criterion \citep[BIC;][]{Schwartz:1978, Fraley:Raftery:1998} and the integrated complete-data likelihood criterion \citep[ICL;][]{Biernacki:Celeux:Govaert:2000}.
When the goal is density estimation, \citet{Roeder:Wasserman:1997} showed that the GMDE model selected using BIC is a consistent estimator of the true density. 
If only the order of the mixture is needed, formal hypothesis testing can also be pursued by likelihood ratio test (LRT). However, standard regularity conditions do not hold for the null distribution of the LRT statistic to have its usual chi-squared distribution \citep[Chap.~6]{McLachlan:Peel:2000}, and significance must be assessed by resampling approaches. For a recent review see \citet{McLachlan:Rathnayake:2014}, and for an implementation in the \code{mclust} software see \citet{Scrucca:etal:2016}.

\section{Methodology}

\subsection{Gaussian mixture density estimation for variables with bounded support}

Let $\x$ be a $p$-variate random vector from a distribution with density $f$ having bounded support $\Space_\XSpace \subset \Real^p$, and $\{t(\x; \lambdab); \lambdab \in \Lambdab\}$ be some family of continuous monotonic transformations that map $\Space_\XSpace$ to an unbounded $p$-dimensional support. Then, we can write $\y = t(\x; \lambdab)$ as the transformed set of variables with density $h$ having unbounded support $\Space_\YSpace$.

Suppose that the density of the transformed data can be expressed through a Gaussian finite mixture density of the form
\begin{equation}
h(\y; \Psib) = \sum_{g=1}^G \pi_g \phi(\y; \mub_g, \Sigmab_g).
\label{eq:denstransf}
\end{equation}
Then, by the continuous change of variable theorem, the density of the untransformed data can be expressed as
\begin{equation*}
f(\x; \Psib, \lambdab) = h(t(\x;\lambdab)) \cdot |\J(t(\x;\lambdab))|,
\end{equation*}
where $\J(t(\x;\lambdab))$ is the Jacobian of the transformation, i.e. the determinant of the matrix of partial derivatives.

\subsection{Range-power transformation for variables with bounded support}

\subsubsection{Lower bound case}
\label{sec:lbound}

Suppose $x$ is a univariate random variable with lower bounded support $\Space_\XSpace  \equiv (l,\infty)$, where $l > -\infty$, and density $f(x)$. Consider a preliminary range transformation defined as $x \mapsto (x - l)$, which maps $\Space_\XSpace \to \Real^{+}$.
Let $\{t(x; \lambda \in \Lambda)\}$ be a continuous monotonic transformation. Based on the well-known Box-Cox transformation \citep{Box:Cox:1964}, we consider the following \emph{range-power} transformation
\begin{equation}
t(x; \lambda) = 
\begin{cases}
\dfrac{(x-l)^{\lambda} - 1}{\lambda} & 
\quad\text{if}\; \lambda \ne 0 \\[1ex]
\log(x-l)                            & 
\quad\text{if}\; \lambda = 0,
\end{cases}
\label{eq:lrangepower}
\end{equation}
which has continuous first derivative equal to 
$t'(x; \lambda) = (x-l)^{\lambda-1}$ for any $\lambda \in \Lambda$.

The original Box-Cox power transformation method is restricted to the univariate case, but it can be extended also to the multivariate case as described in \citet{Velilla:1993}. 
However, further development of the multivariate case $\x = (x_1, \ldots, x_p)\T$ can greatly simplified by working in a coordinate-wise fashion. Thus, in this paper we propose the use of the range-power transformation in \eqref{eq:lrangepower} for each dimension separately. 

\subsubsection{Lower and upper bound case}
\label{sec:lubound}

Suppose now that $x$ is a univariate random variable with bounded support $\Space_X \equiv (l,u)$, where $-\infty < l < u < +\infty$. 
Consider the preliminary range transformation $x \mapsto (x - l)/(u - x)$ which maps  $\Space_X \to \Real^{+}$. As in the previous case, adopting a \emph{range-power} transformation we can write
\begin{equation}
t(x; \lambda) = 
\begin{cases}
\dfrac{ \left( \dfrac{x-l}{u-x} \right)^{\lambda} - 1}{\lambda}
  & \quad\text{if}\; \lambda \ne 0 \\[2ex]
\log \left( \dfrac{x-l}{u-x} \right)
  & \quad\text{if}\; \lambda = 0,
\end{cases}
\label{eq:lurangepower}
\end{equation}
with continuous first derivative given by
\begin{equation*}
t'(x; \lambda) = 
\begin{cases}
\left( \dfrac{x-l}{u-x} \right)^{\lambda-1} \dfrac{u-l}{(u-x)^2} 
  & \text{if}\; \lambda \ne 0 \\[2ex]
\dfrac{1}{x-l} + \dfrac{1}{u-x}
  & \text{if}\; \lambda = 0.
\end{cases}
\end{equation*}

Following the approach discussed in Section~\ref{sec:lbound}, the multivariate case can be tackled by working in a coordinate-wise fashion, hence applying the range-power transformation in \eqref{eq:lurangepower} to each variable separately.

\subsection{Estimation}
\label{sec:estimation}

Maximum likelihood estimation can be pursued via the EM algorithm under the assumption that the density on the transformed scale can be expressed as in \eqref{eq:denstransf}, with $\y = t(\x; \lambdab)$ the vector of range-power transformed variables according to \eqref{eq:lrangepower} or \eqref{eq:lurangepower}. 
If the previous assumption holds, then the density function on the original scale is given by 
\begin{equation}
f(\x; \Psib, \lambdab) = 
\sum_{g=1}^G \pi_g \phi( t(\x; \lambdab); \mub_g, \Sigmab_g) \cdot
                   |\J(t(\x; \lambdab)|,
\label{eq:densuntrans}
\end{equation}
where 
$t(\x; \lambdab) = (t(x_1; \lambda_1), \ldots, t(x_p; \lambda_p))\T$ 
and  
$\J(t(\x; \lambdab))$ is the Jacobian of the transformation. 
Note that as consequence of the coordinate independent approach to multivariate range-power transformation, the matrix of first derivatives is diagonal, so the Jacobian reduces to 
\begin{equation*}
\J(t(\x; \lambdab)) = 
\det\left[ \frac{\partial t(\x; \lambdab)}{\partial \x} \right] = 
\prod_{j=1}^p \frac{\partial t(x_j; \lambda_j)}{\partial x_j}.
\end{equation*}

The conditional expectation of the complete log-likelihood given the observed data can be expressed as 
\begin{equation*}
Q(\Psib; \Psib^{(m)}) = \sum_{i=1}^n \sum_{g=1}^G \hat{z}_{ig}^{(m)} 
\left\{ \log\pi_g + 
        \log\phi(t(\x_i; \lambdab); \mub_g, \Sigmab) +
        \log|\J(t(\x_i; \lambdab))|
\right\},
\end{equation*}
where $\hat{z}_{ig}^{(m)} = \Exp(I(z_i=g)|\x_i, \Psib^{(m)})$. 
Therefore, in the E-step the posterior probabilities are updated using
\begin{equation*}
\hat{z}_{ig}^{(m+1)} = 
  \dfrac{\hat{\pi}_g^{(m)} 
         \phi\left(t(\x_i; \hat{\lambdab}^{(m)}); 
                   \hat{\mub}_g^{(m)}, \hat{\Sigmab}_g^{(m)}
             \right)}
        {\sum_{k=1}^G \hat{\pi}_k^{(m)} 
         \phi\left(t(\x_i; \hat{\lambdab}^{(m)});
                   \hat{\mub}_k^{(m)}, \hat{\Sigmab}_k^{(m)}
             \right)}.
\end{equation*} 

In the M-step the parameters $(\Psib, \lambdab)$ are updated by maximising the $Q$-function given the previous values of the parameters and the updated posterior probabilities. This can be done in two steps. 
In the first step an updated value $\hat{\lambdab}^{(m+1)}$ is computed by numerically maximising the Q-function with respect to $\lambdab$ because no closed-form expression is available. To this goal, a Newton-type numerical optimisation algorithm can be used. In our implementation we used the \code{L-BFGS-B} method of \citet{Byrd:etal:1995} available in the \code{optim()} function for the \textsf{R} statistical software.
The remaining parameters are then obtained as in standard EM algorithm but accounting for the updated transformation parameters $\hat{\lambdab}^{(m+1)}$, i.e.
\begin{equation*}
\hat{\pi}_g^{(m+1)} = \dfrac{\sum_{i=1}^n \hat{z}_{ig}^{(m+1)}}{n}
\qquad\text{and}\qquad
\hat{\mub}_g^{(m+1)} = 
  \dfrac{\sum_{i=1}^n \hat{z}_{ig}^{(m+1)} t(\x_i; \hat{\lambdab}^{(m+1)})}
        {\sum_{i=1}^n \hat{z}_{ig}^{(m+1)}}.
\end{equation*}
Again, the update formula for the covariance matrix depends upon the assumed eigendecomposition model. In the most general case of an unconstrained covariance matrix, i.e. the VVV model, we have 
\begin{equation*}
\hat{\Sigmab}_g^{(m+1)} = 
  \dfrac{\sum_{i=1}^n \hat{z}_{ig}^{(m+1)} 
         \left(t(\x_i; \hat{\lambdab}^{(m+1)}) - \hat{\mub}_{g}^{(m+1)}\right)
         \left(t(\x_i; \hat{\lambdab}^{(m+1)}) - \hat{\mub}_{g}^{(m+1)}\right)\T}
        {\sum_{i=1}^n \hat{z}_{ig}^{(m+1)}}.
\end{equation*}

Initialisation of the above EM algorithm is obtained by first estimating the optimal marginal transformations, then using the final classification from a $k$-means algorithm on the range-power transformed variables. This initial partition of data points is used to start the algorithm from the M-step. 
Finally, the EM algorithm is stopped when the log-likelihood improvement falls below a specified tolerance value or a maximum number of iterations is reached.

\section{Simulation studies}

In this section we present some simulation studies designed to compare the proposed GMDEB approach to some density estimators for bounded variables discussed in the literature. 
The comparison is based on the integrate squared error (ISE):
\begin{equation*}
\ISE(\hat{f}) = \int \left[\hat{f}(x) - f(x) \right]^2 dx,
\end{equation*}
where $f$ is the unknown true density and $\hat{f}$ is its estimate based on a random sample of $n$ observations. Thus, the $\ISE$ is a measure of discrepancy between the true and the estimated density based on a squared loss criterion. It is equal to 0 when the estimated density perfectly coincides with the true density, and increases as the differences between the two densities get larger. For more details see \citet[][Sec. 2.3]{Scott:2009}. 
In the following sections the ISE is computed via numerical integration.

\subsection{Distributions with lower bound}

\begin{figure}[htb]
\centering
\includegraphics[width=\textwidth]{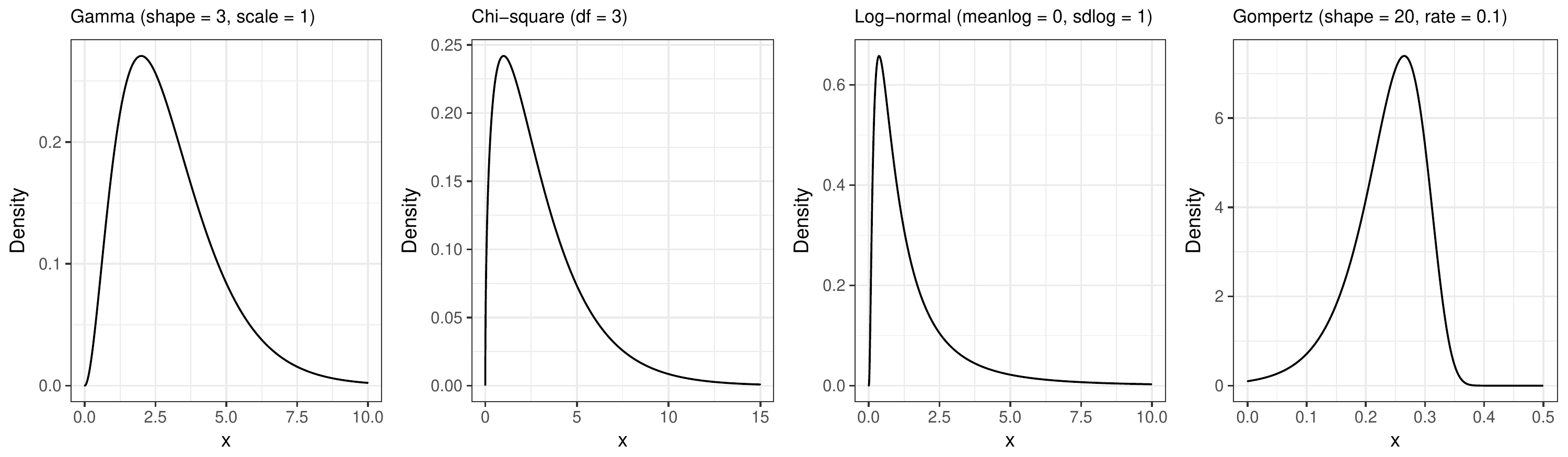}
\caption{\small Univariate densities with lower bound considered in the simulation study.}
\label{fig1:lbound}
\end{figure}

The univariate densities with lower bound support considered in this study are shown in Figure~\ref{fig1:lbound}. They are all bounded at zero with different degrees of skewness, positive for the first three densities and negative for the last one.

The proposed method (\code{GMDEB}) is compared with the following density estimators:
\begin{itemize}
\item \code{simpleKern} which refers to the simple boundary correction method proposed by \citet{Jones:1993} which is equivalent to a kernel weighted local linear fitting near the boundary;
\item \code{reflectKern} which indicates the reflection method of \citet{Schuster:1985} which amounts to reflect the observed data points at the origin, then a density estimate is obtained using this augmented dataset with a simple correction to ensure that integrate to one;
\item \code{cutnormKern} which is the cut and normalisation method of \citet{Gasser:Muller:1979} where the kernel is truncated at the boundary and re-normalised to unity;
\item \code{logtransKern} which is the method proposed by \citet{Marron:Ruppert:1994} which fits a kernel density estimator on the log-scale and then back-transforms the result with an explicit normalisation step;
\item \code{logSpline} which estimates a density using cubic splines to approximate the log-density using knots located as described in \citet{Stone:etal:1997};
\item \code{GaMixDE} which estimates a density by fitting a mixture of Gamma densities;
\item \code{GMDE} which is the standard density estimate from GMMs with no boundary correction.
\end{itemize}

The first four methods mentioned above are implemented in the \code{evmix} R package \citep{Rpkg:evmix, Hu:Scarrott:2018}, whereas the \code{logSpline} estimator is available in the \code{logspline} R package \citep{Rpkg:logspline}. 
For \code{GaMixDE} the code is available in the R package \code{mixtools} \citep{Benaglia:etal:2009, Rpkg:mixtools}, whereas \code{GMDE} is obtained from the \code{mclust} R package \citep{Rpkg:mclust}. In the last two cases the number of mixture components is selected using the BIC criterion.

\begin{figure}[htb]
\centering
\includegraphics[width=0.49\textwidth]{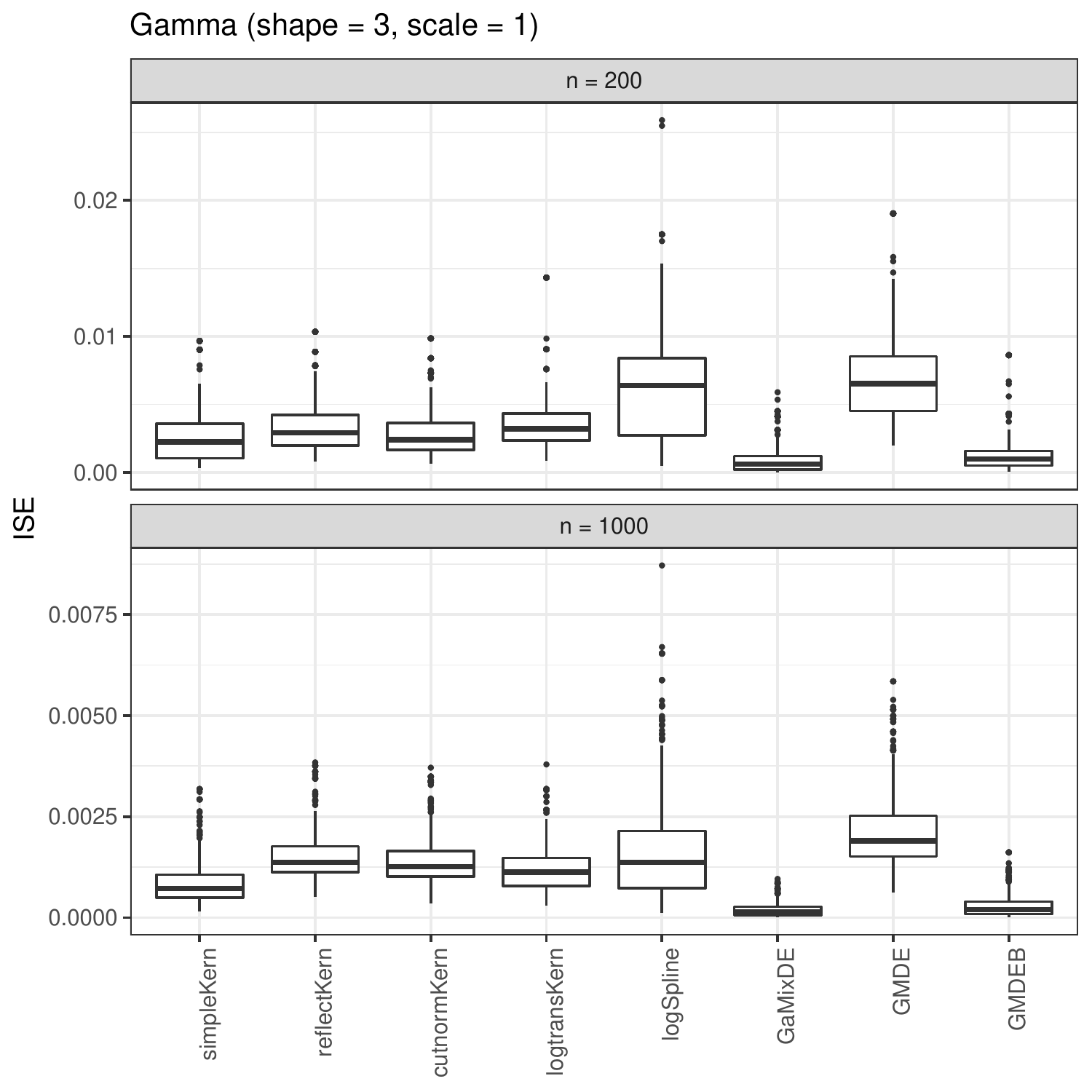}
\includegraphics[width=0.49\textwidth]{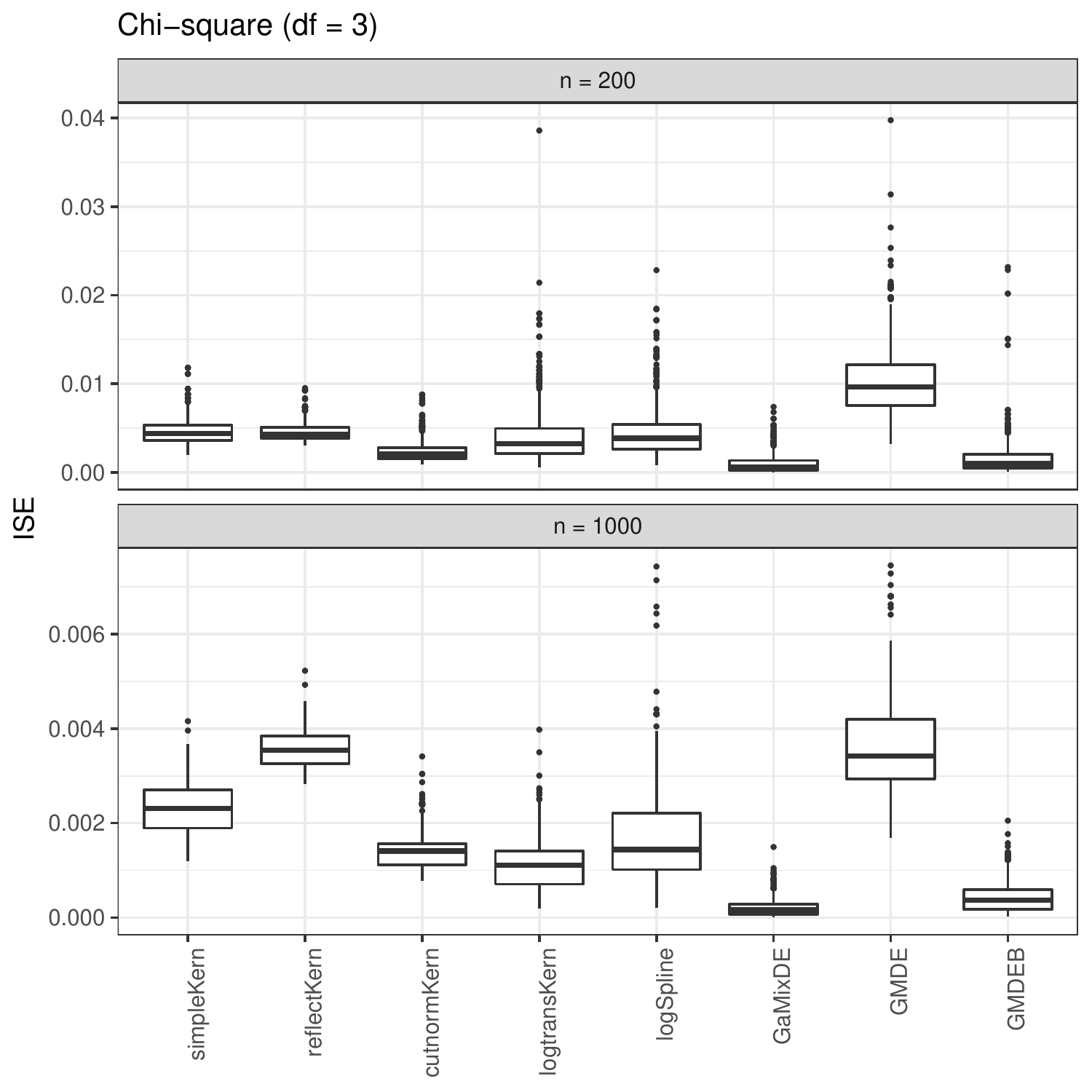} \\
\includegraphics[width=0.49\textwidth]{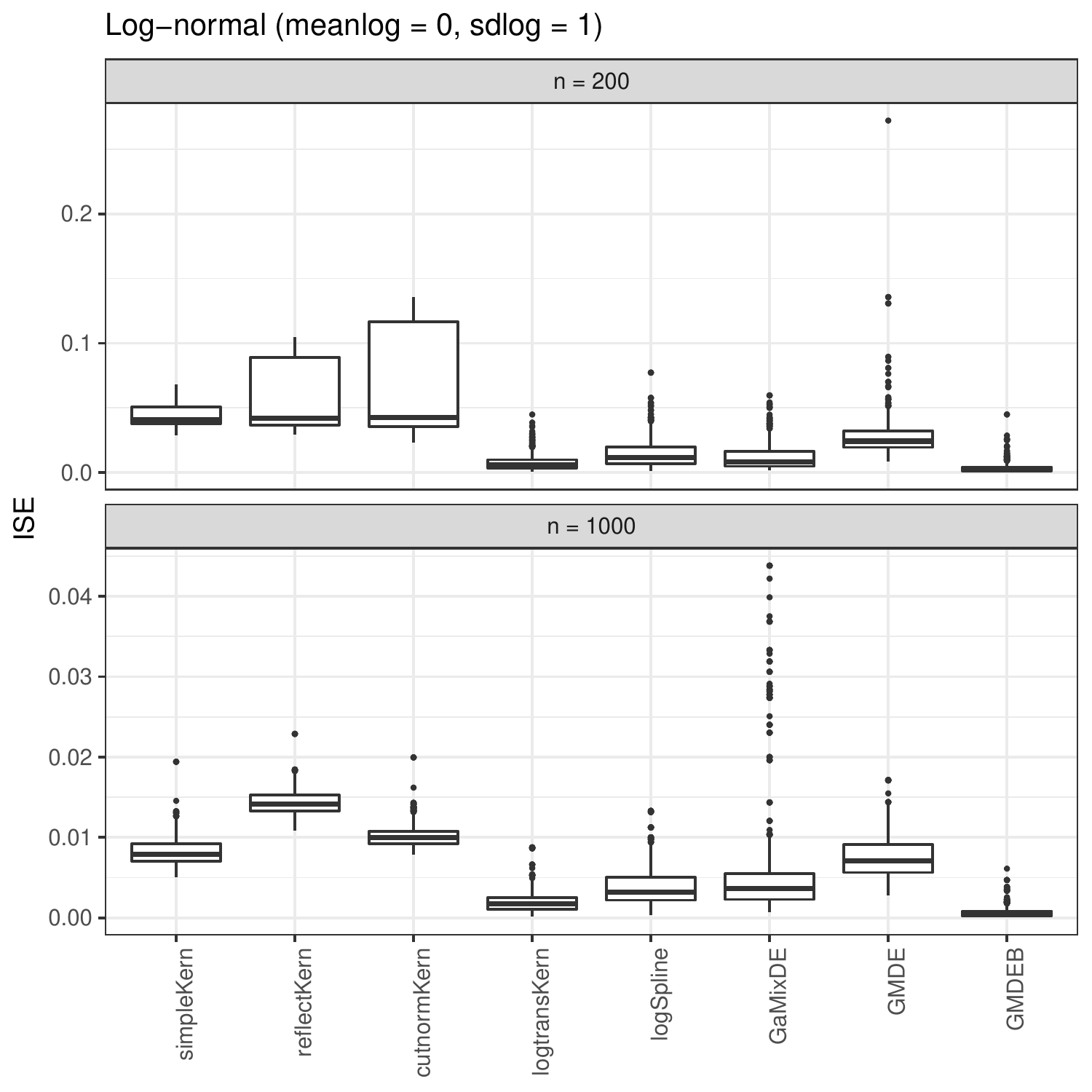}
\includegraphics[width=0.49\textwidth]{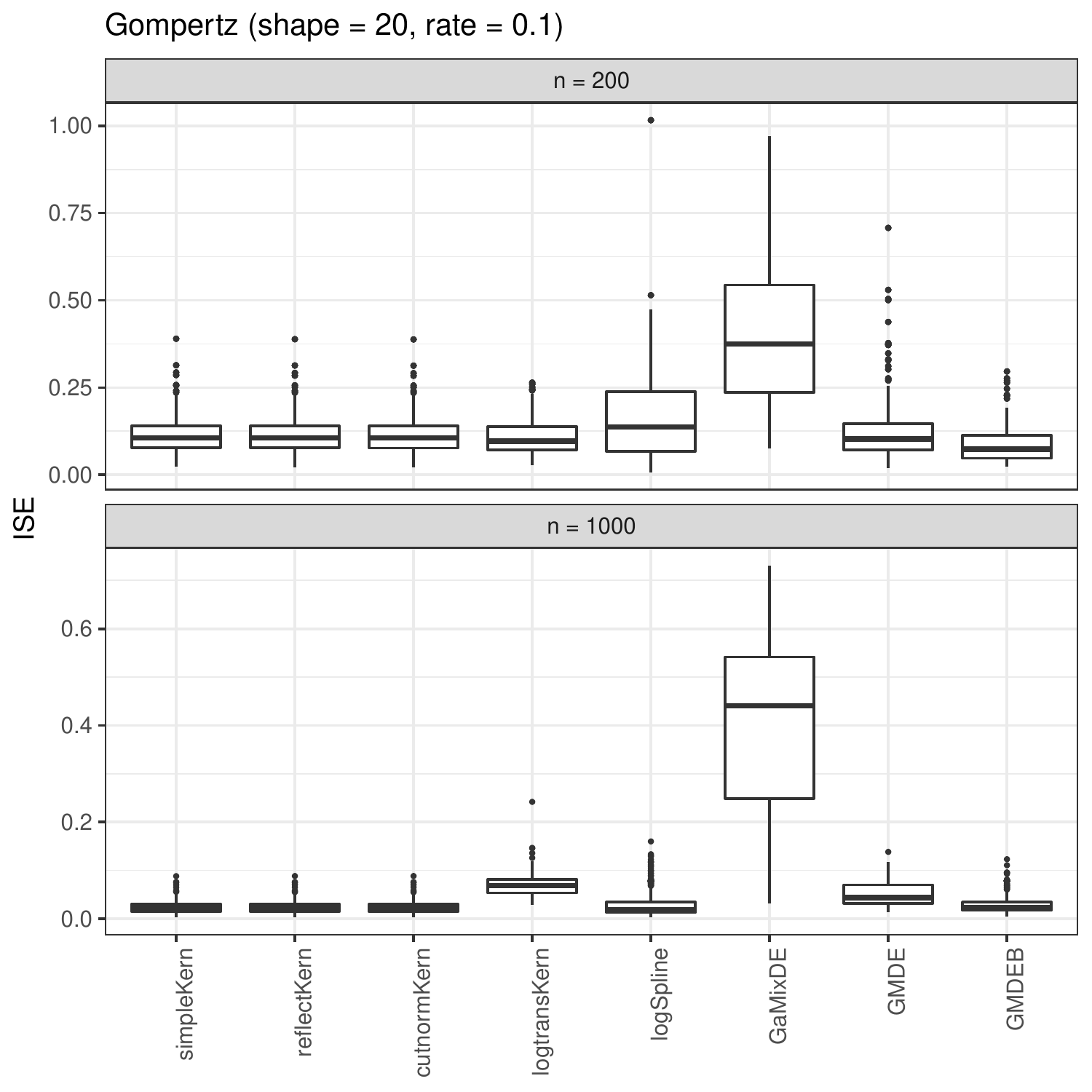}
\caption{\small Boxplots of ISE distribution from 1000 replications of the simulation study for the selected univariate densities with lower bound.}
\label{fig2:lbound}
\end{figure}

Figure~\ref{fig2:lbound} graphically summarises the simulation results obtained on 1000 replications. Overall the GMDEB estimator appears to be able to approximate the true density better than the other KDE methods with boundary correction, in particular when the sample size is small. The proposed approach also shows less variability, which decreases as the sample size increases. Clearly the GMDEB approach appears to be inferior to the Gamma mixture density estimator, although not by much, when the true density belongs to the Gamma family of distributions (i.e. in the first two cases), but it is better in the last two cases, in particular for the left-skewed Gompertz distribution.

\clearpage

\subsection{Distributions with lower and upper bounds}

\begin{figure}[htb]
\centering
\includegraphics[width=\textwidth]{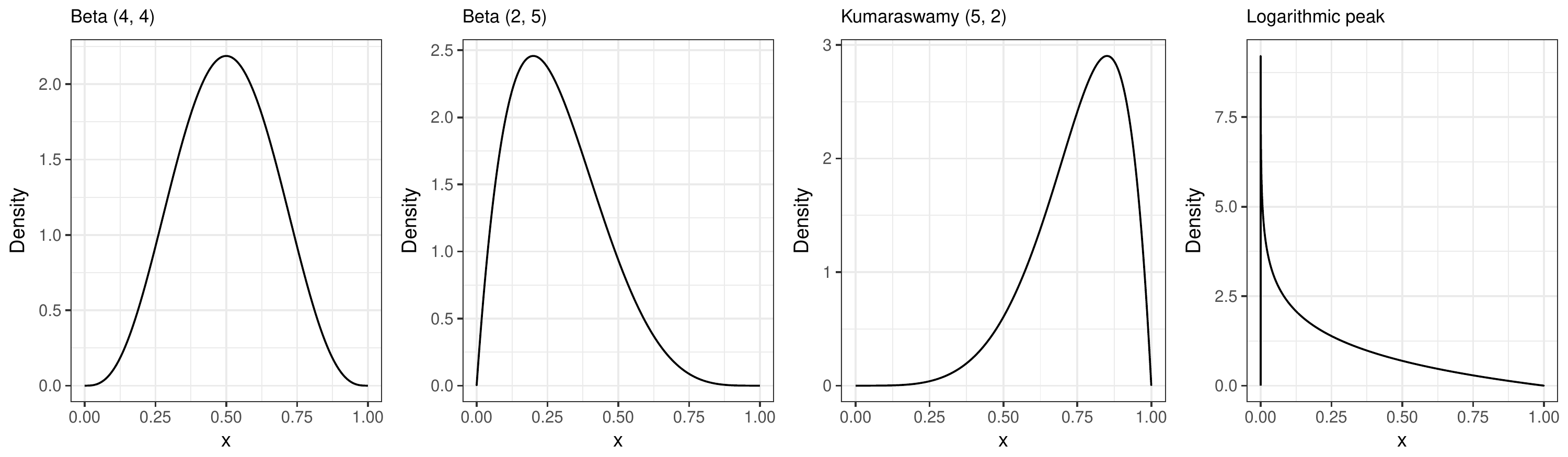}
\caption{\small Univariate densities with lower and upper bounds considered in the simulation study.}
\label{fig1:lubound}
\end{figure}

For the case of univariate densities with both lower and upper bounds support, a list of distributions considered in the simulation study is shown in Figure~\ref{fig1:lubound}. The first two settings involve the Beta distribution with parameters selected to produce a symmetric case and a skewed case. The last two settings consider two further asymmetric cases, the Kumaraswamy distribution with density $f(x) = \alpha\beta x^{\alpha-1}{(1-x^{\alpha})}^{\beta-1}$ on $[0,1]$, and the logarithmic peak distribution with density $f(x) = -\log(x)$ on $(0, 1)$. 

The cases described above should provide a broad spectrum of examples for comparing different estimators. To this goal the proposed method (\code{GMDEB}) is compared with the following density estimators:
\begin{itemize}
\item \code{beta1Kern}, \code{beta2Kern} which use the Beta and modified Beta kernels proposed by \citet{Chen:1999} followed by a renormalisation to ensure a proper density; 
\item \code{copulaKern} which uses the bivariate Gaussian copula based kernels of \citet{Jones:Henderson:2007};
\item \code{logSpline} which fits a density using cubic splines to approximate the log-density using knots located as described in \citet{Stone:etal:1997};
\item \code{BeMixDE} which estimates the density by fitting a mixture of Beta distributions 
\end{itemize}

The first two estimators are available in the R package \code{evmix} \citep{Rpkg:evmix, Hu:Scarrott:2018}. The \code{logSpline} estimator is available in the \code{logspline} R package \citep{Rpkg:logspline}. 
For the \code{BeMixDE} the \code{betareg} R package \citep{Grun:Kosmidis:Zeileis:2012} is used with the number of mixture components selected using BIC.

\begin{figure}[htb]
\centering
\includegraphics[width=0.49\textwidth]{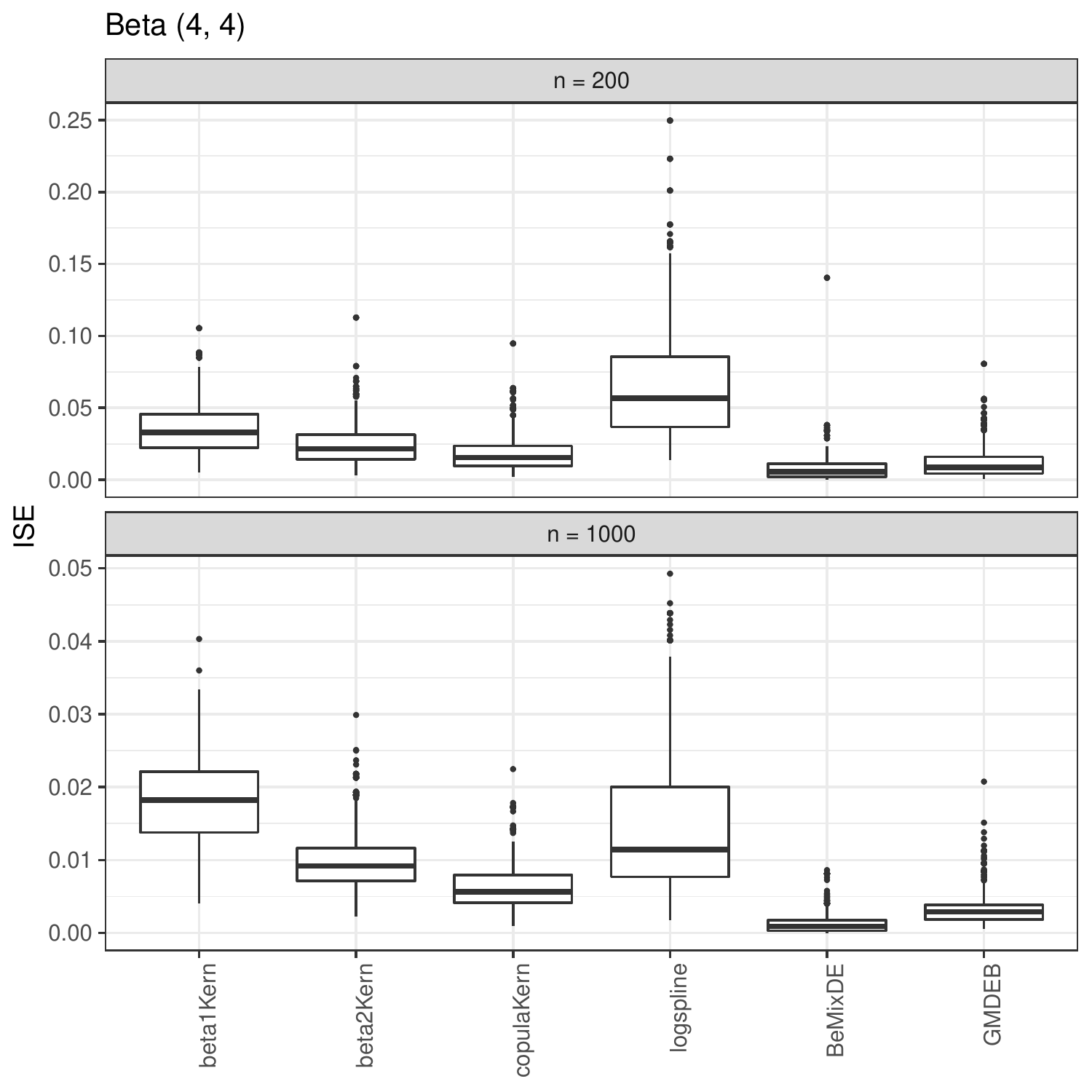}
\includegraphics[width=0.49\textwidth]{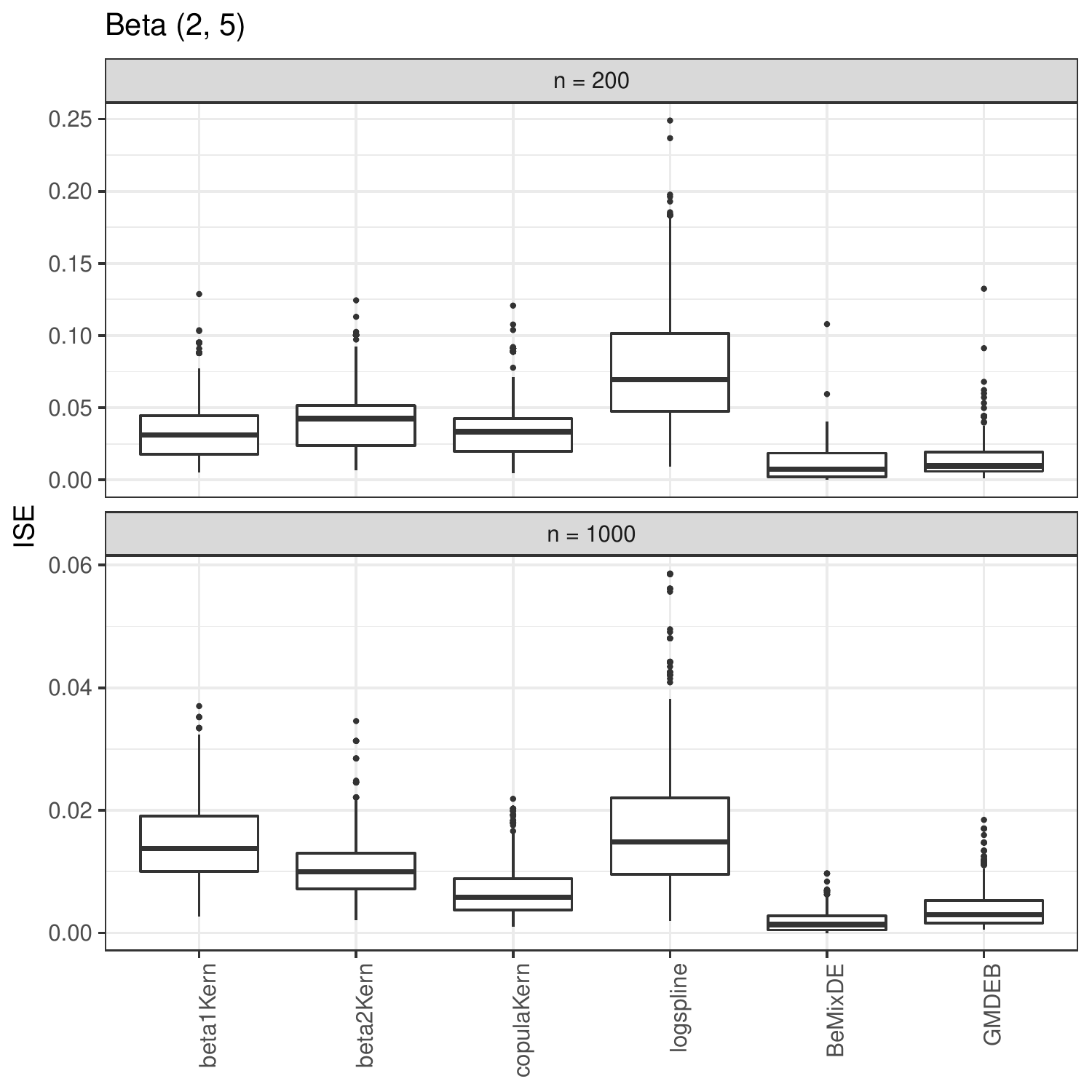} \\
\includegraphics[width=0.49\textwidth]{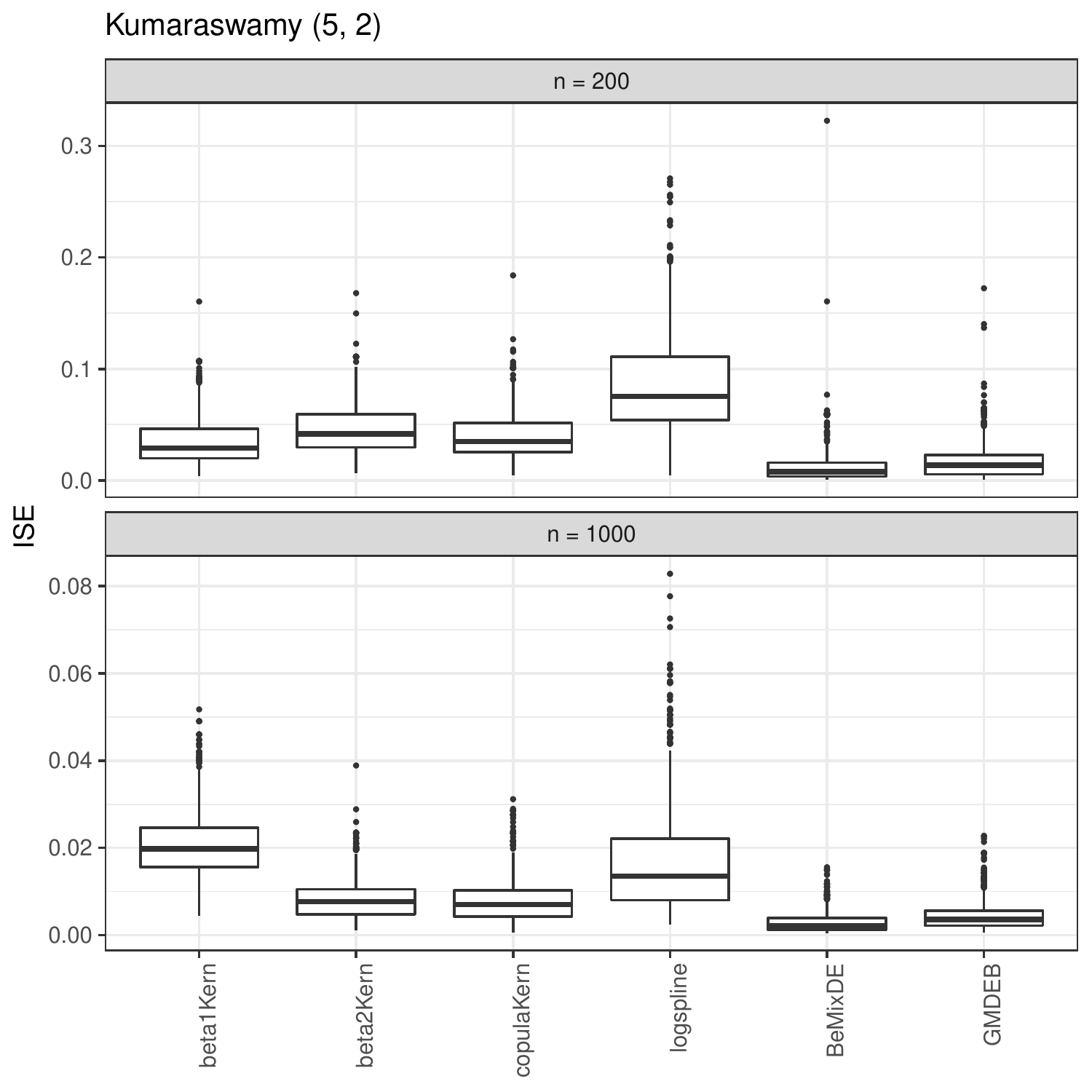} 
\includegraphics[width=0.49\textwidth]{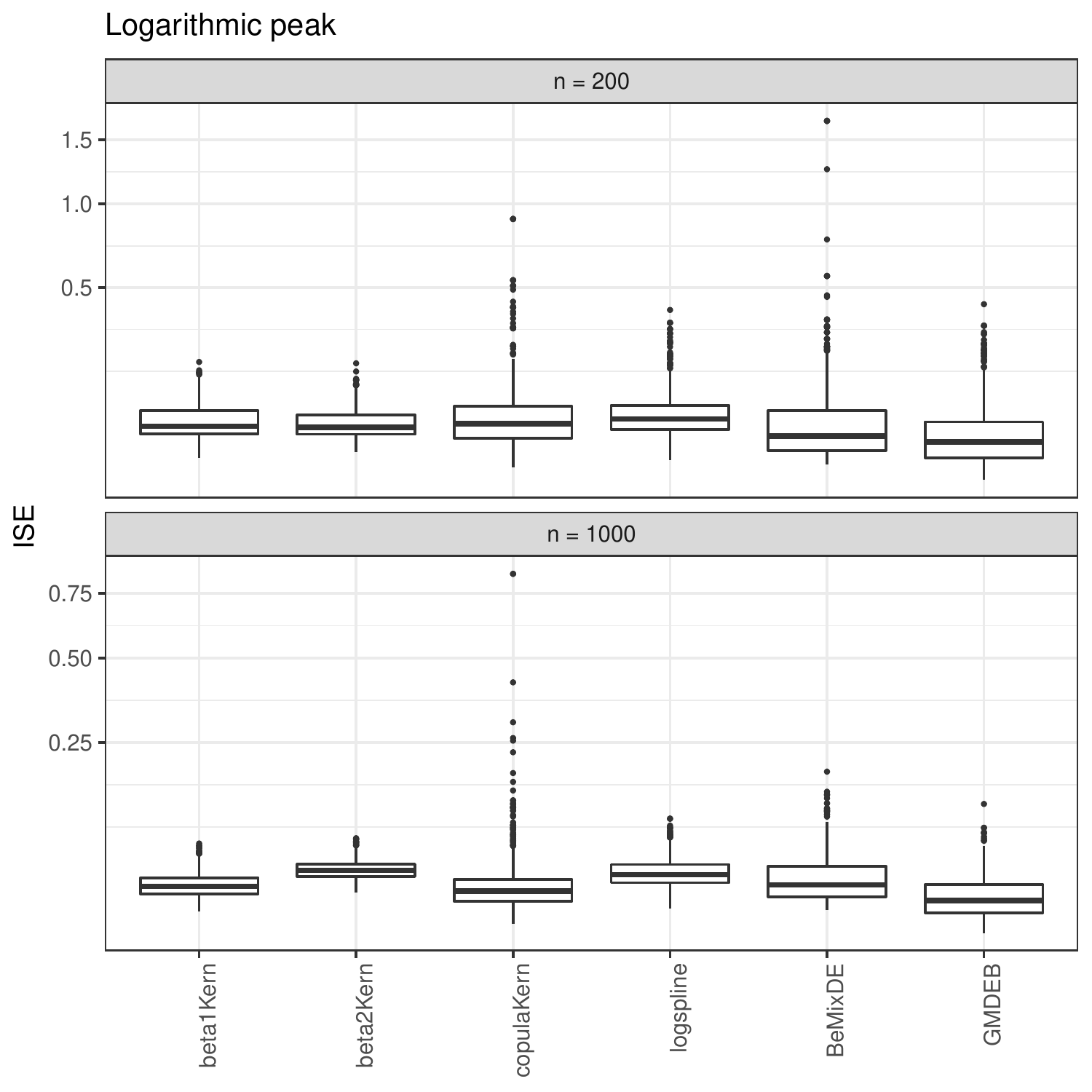} 
\caption{\small Boxplots of ISE distribution from 1000 replications of the simulation study for the selected univariate densities with lower and upper bounds.}
\label{fig2:lubound}
\end{figure}

Figure~\ref{fig2:lubound} reports the simulation results obtained on 1000 replications. By looking at the boxplots the GMDEB approach appears to be more accurate than the other non-parametric density estimators. Furthermore, its accuracy is slightly lower than the Beta mixture density estimator in the first three cases (which, however, are all cases related to the Beta distribution), but is better in the last case. Overall, GMDEB seems to provide robust reliable density estimates when both lower and upper bounds are present.

\subsection{A note on computing time}

A major concern might be the computational effort required by the estimation of the $\lambdab$ parameter through numerical optimisation within the EM algorithm. 
To investigate the runtime of the proposed procedure, we designed a small simulation study where we generated data from a Chi-square distribution with 3 degrees of freedom for sample sizes $n = \{200, 1000, 10000\}$. A multivariate case was also investigated by considering a 10-dimensional variable with independent marginals drawn as in the univariate case. 
We estimated the density of GMDEB by both estimating the lambda parameter, and by fixing it at the corresponding MLE value. Furthermore, we executed the algorithm both sequentially and in parallel (over the mixture components and covariance parameterisations).
Experiments were carried out on an iMac with 4 cores i5 Intel CPU running at 2.8 GHz and with 16GB of RAM.

\begin{figure}[htb]
\centering
\includegraphics[width=0.7\textwidth]{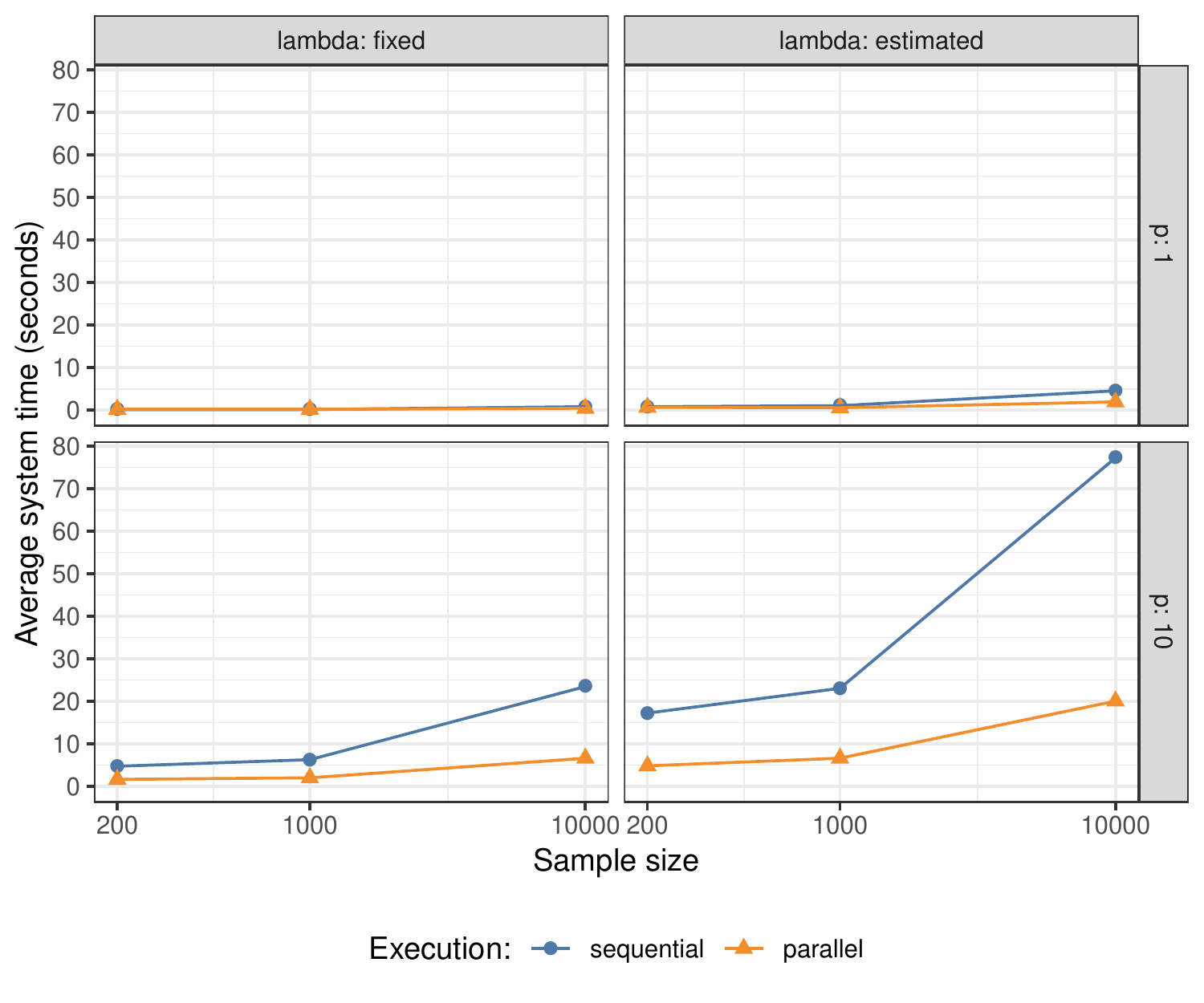}
\caption{\small Average runtimes for different sample sizes (expressed in $\log_{10}$-scale) obtained by considering the transformation parameter either fixed or to be estimated, by running the GMDEB algorithm sequentially or in parallel, and for number of variables 1 and 10.}
\label{fig:systime}
\end{figure}

Figure~\ref{fig:systime} shows the results averaged over 100 replications of the experiments. Clearly, in the univariate case the effect of estimating the transformation parameter is negligible. On the contrary, the effect is visible in the multivariate case, but a considerable speedup can be achieved by parallelisation. The worst case, i.e. transformation parameter to be estimated sequentially with a sample of size 10000 on 10 dimensions, required on average just over 1 minute. 

\section{Real data analyses}

\subsection{Acidity data}

This dataset provides the values of an acidity index (acid-neutralizing capacity, \code{ANC}) measured in a sample of 155 lakes in North-Central Wisconsin. Several authors have previously analysed the data using a mixture of Gaussian distributions on the log-scale \citep{Crawford:etal:1992, Crawford:1994, Richardson:Green:1997, McLachlan:Peel:2000}. 
On the contrary, we analyse the data in the original scale because the proposed method automatically selects the ``optimal'' transformation and takes into account the implicit lower bound of the index that can not assume negative values. 

\begin{figure}[htb]
\centering
\includegraphics[width=0.49\textwidth]{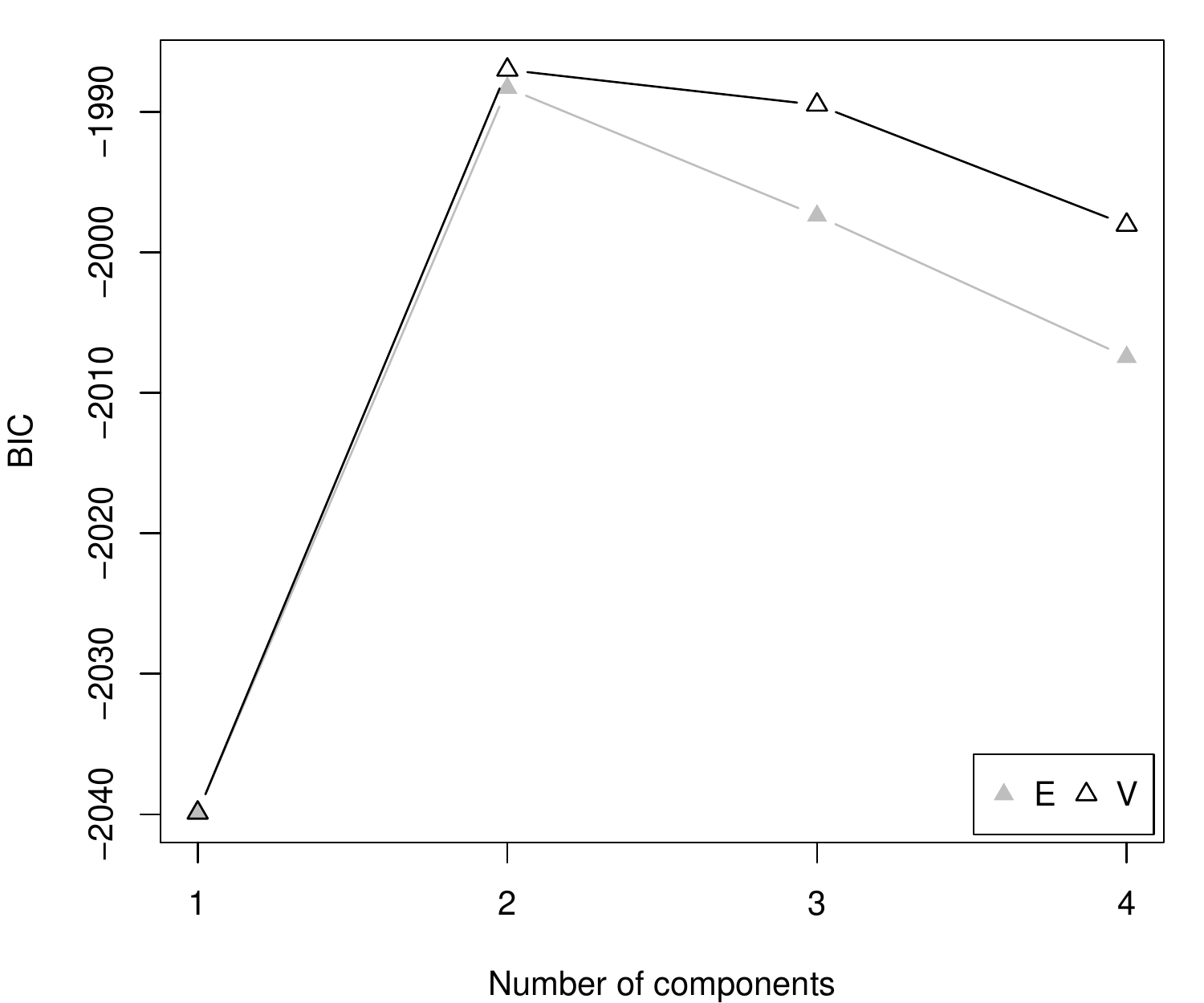}
\includegraphics[width=0.49\textwidth]{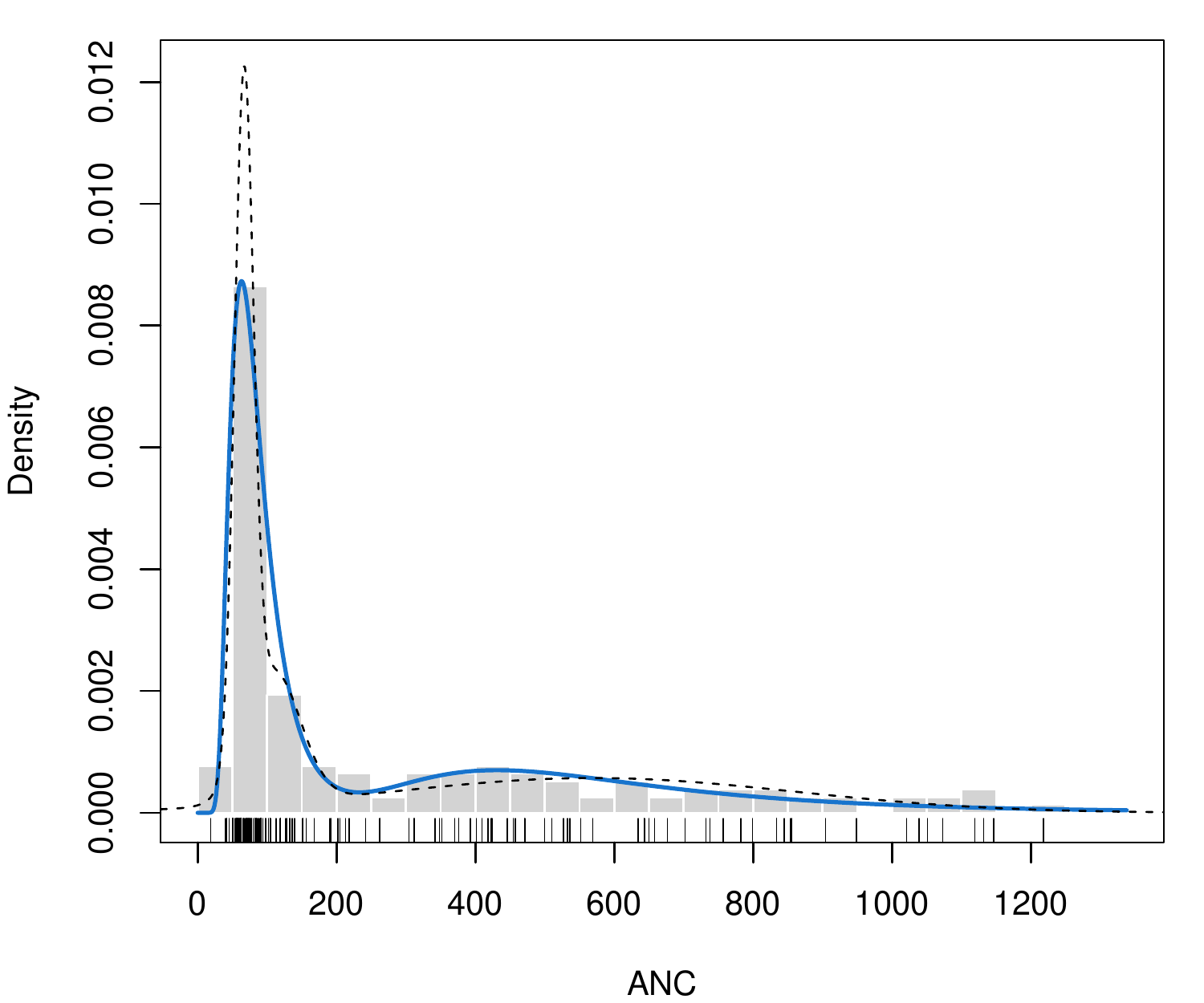}
\caption{\small Plot of BIC values for different number of mixture components and within-component variances (left panel). Histogram of acidity data with GMDEB (blue thick line) and GMM (black dashed line) density estimates (right panel).}
\label{fig1:acidity}
\end{figure}

\begin{figure}[htb]
\centering
\includegraphics[width=0.49\textwidth]{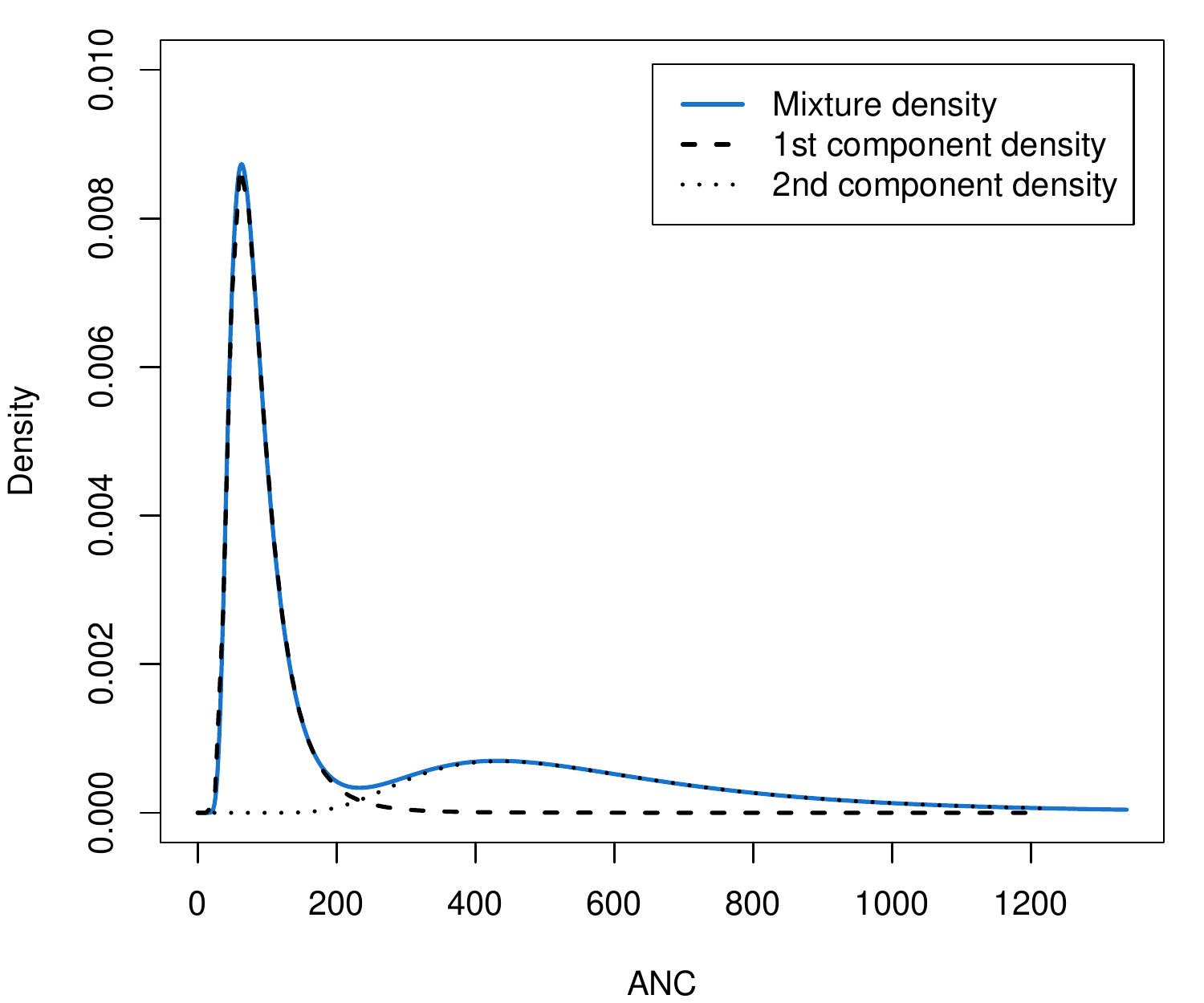}
\includegraphics[width=0.49\textwidth]{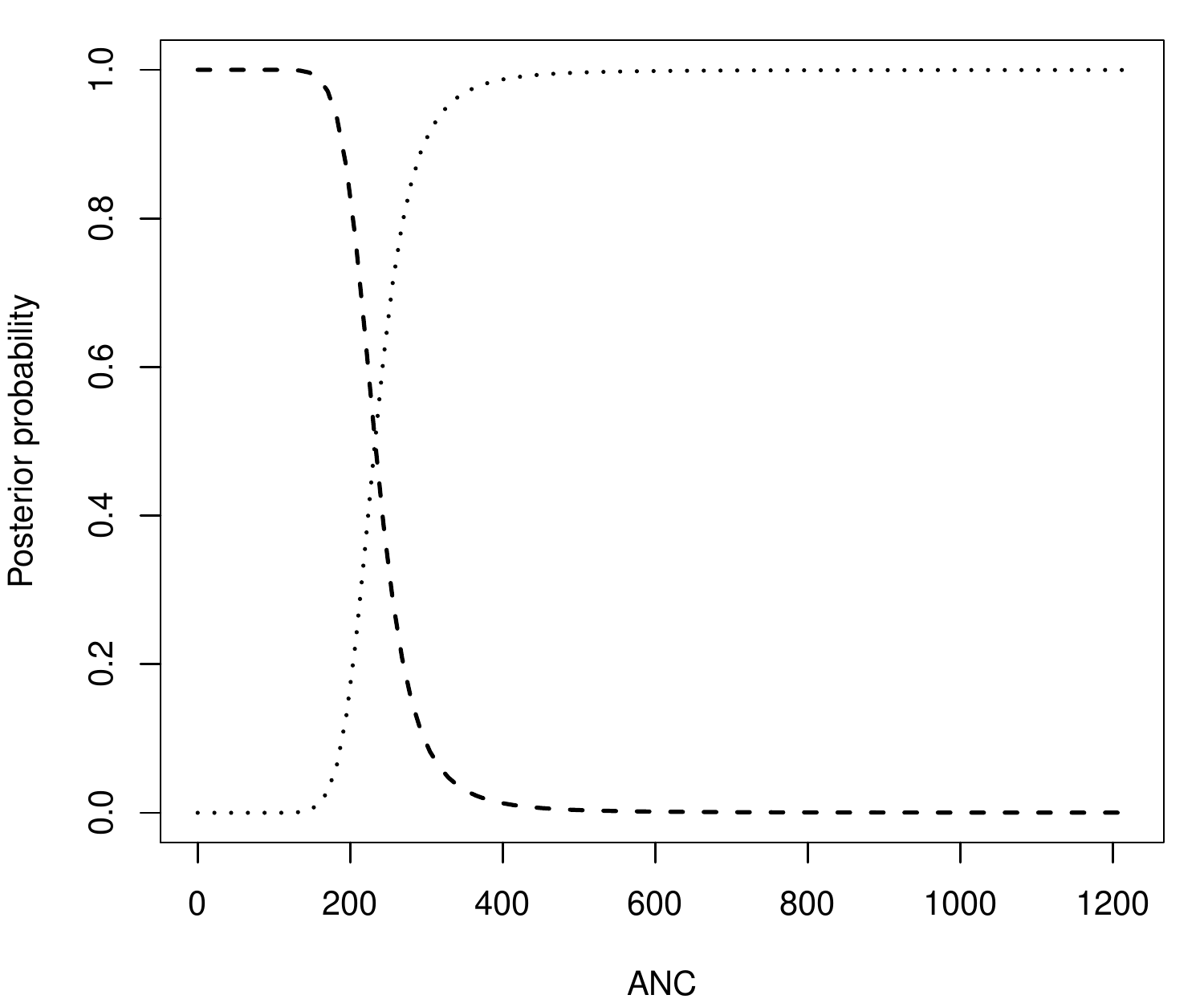}
\caption{\small Plot of estimated mixture density and rescaled component densities (left panel), and corresponding estimated posterior probabilities for the acidity data (right panel).}
\label{fig2:acidity}
\end{figure}

From the left panel of Figure~\ref{fig1:acidity} we can see that according to BIC the best model is the one with two mixture components having different variances (V,2), closely followed by models (E,2) and (V,3). The right panel of Figure~\ref{fig1:acidity} shows the histogram of the data and the density estimated with model (V,2) using the GMDEB approach with transformation parameter $\hat{\lambda}=-0.293$ (blue thick line). 
For comparison we also draw the density estimated by GMM without any boundary correction (black dashed line). 
The density estimated by GMDEB appears to accurately follow the distribution of the data, indicating the presence of two separated skewed distributions having different dispersions, smaller for the component close to the origin and larger for higher values of \code{ANC}. 
This is also confirmed by the graphs in Figure~\ref{fig2:acidity}, which show the component densities scaled by the estimated prior probabilities $\hat{\pi} = (0.6322, 0.3678)$ (left panel) and the estimated posterior probabilities $\hat{z}_{ij}$. Using the standard cut-off value of $0.5$, lakes with \code{ANC} smaller than about 232 are assigned to the first group, otherwise to the second group. This is in agreement with previous findings.

\subsection{Racial data}

\citet{Geenens:2013} presented an analysis on data giving the proportion of white student enrolled in 56 school districts in Nassau County (Long Island, New York), for the 1992--1993 school year. The density estimate for this dataset should only be supported on the $[0,1]$ range. See also \citet[][Sec. 3.2]{Simonoff:1996}.

\begin{figure}[htb]
\centering
\includegraphics[width=\textwidth]{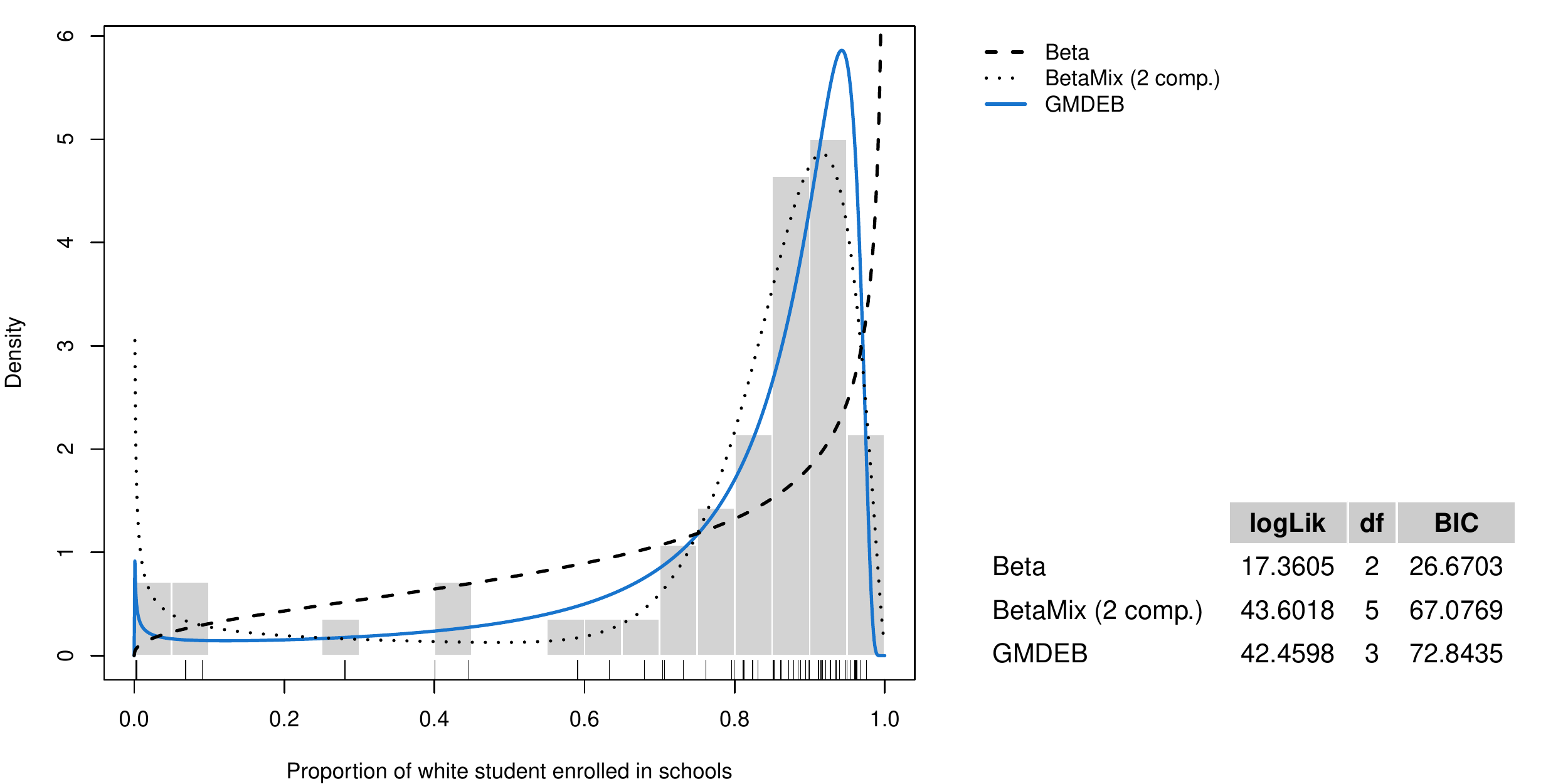}
\caption{\small Density estimates for the racial data obtained using the GMDEB method, the standard Beta distribution, and a two-components Beta mixture. The included table reports the log-likelihood, the number of estimated parameters, and the BIC (larger values are preferred).}
\label{fig1:racial}
\end{figure}

The selected model on the transformed scale is (E,1) with $\hat{\lambda}=0.387$, and the corresponding density estimate on the original scale is shown in Figure~\ref{fig1:racial}. This can be compared graphically with the Beta density and the Beta mixture using two components. Both models were estimated by maximum likelihood using the \code{betareg} R package \citep{Grun:Kosmidis:Zeileis:2012}, with a single component in the first case, and the optimal number of mixture components selected using BIC in the second case.
The single-component Beta density seems to put too much emphasis close to the upper boundary and in the middle values of the distribution, while completely missing the bulk of the data between 70\% and 90\% of white students. On the contrary, the proposed GMDEB approach provides a density estimate which correctly identify the majority of the data with at least 70\% of white students, but also the small peak near the lower boundary containing schools with almost 0\% white students. The two-components Beta mixture density is quite close to that provided by GMDEB, but the latter should be preferred according to BIC (see table in Figure~\ref{fig1:racial}).
These findings largely agree with those reported in \citet[][Fig. 3]{Geenens:2013}.


\subsection{Plasma data}

\begin{figure}[htb]
\centering
\includegraphics[width=0.48\textwidth]{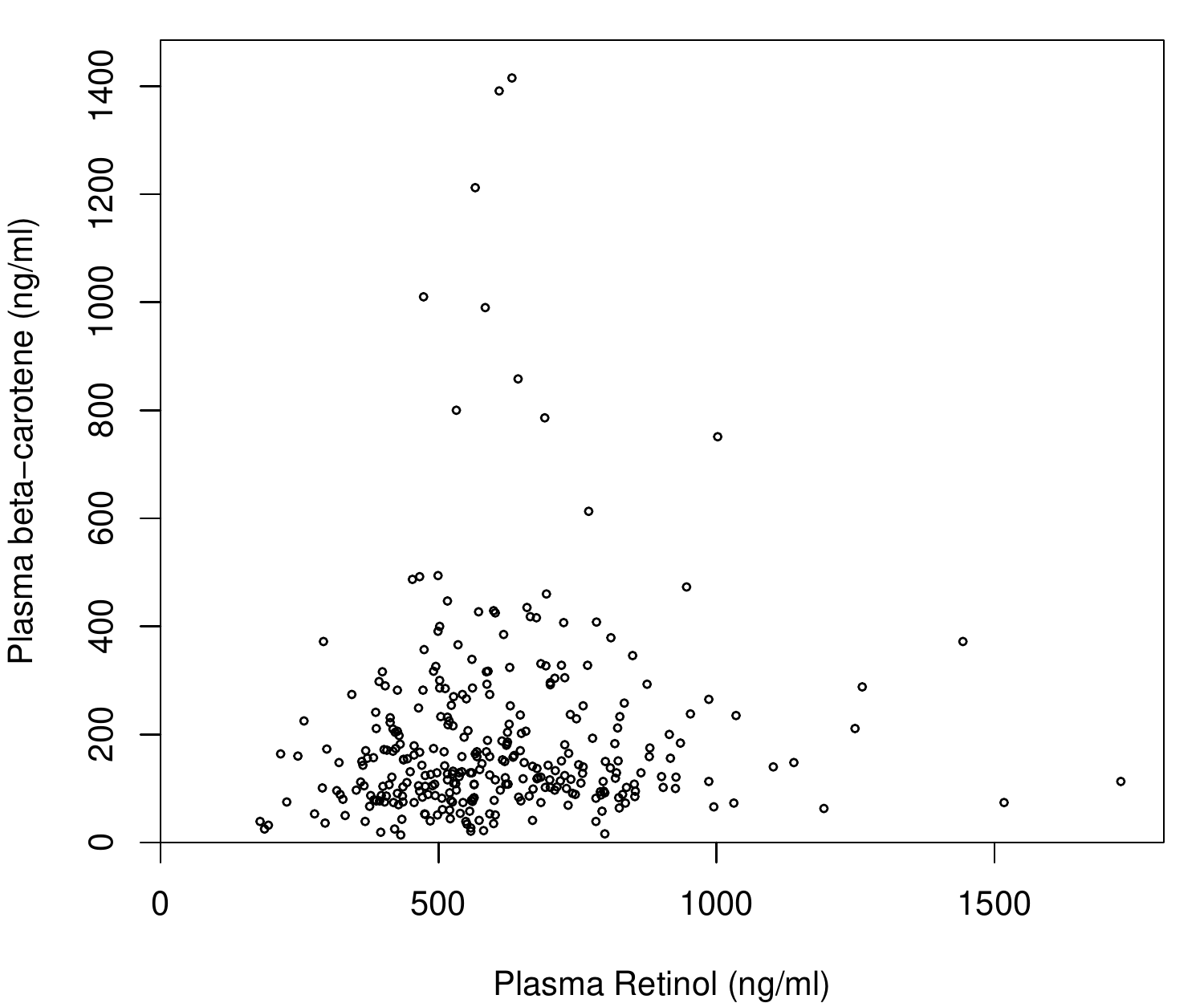}
\includegraphics[width=0.48\textwidth]{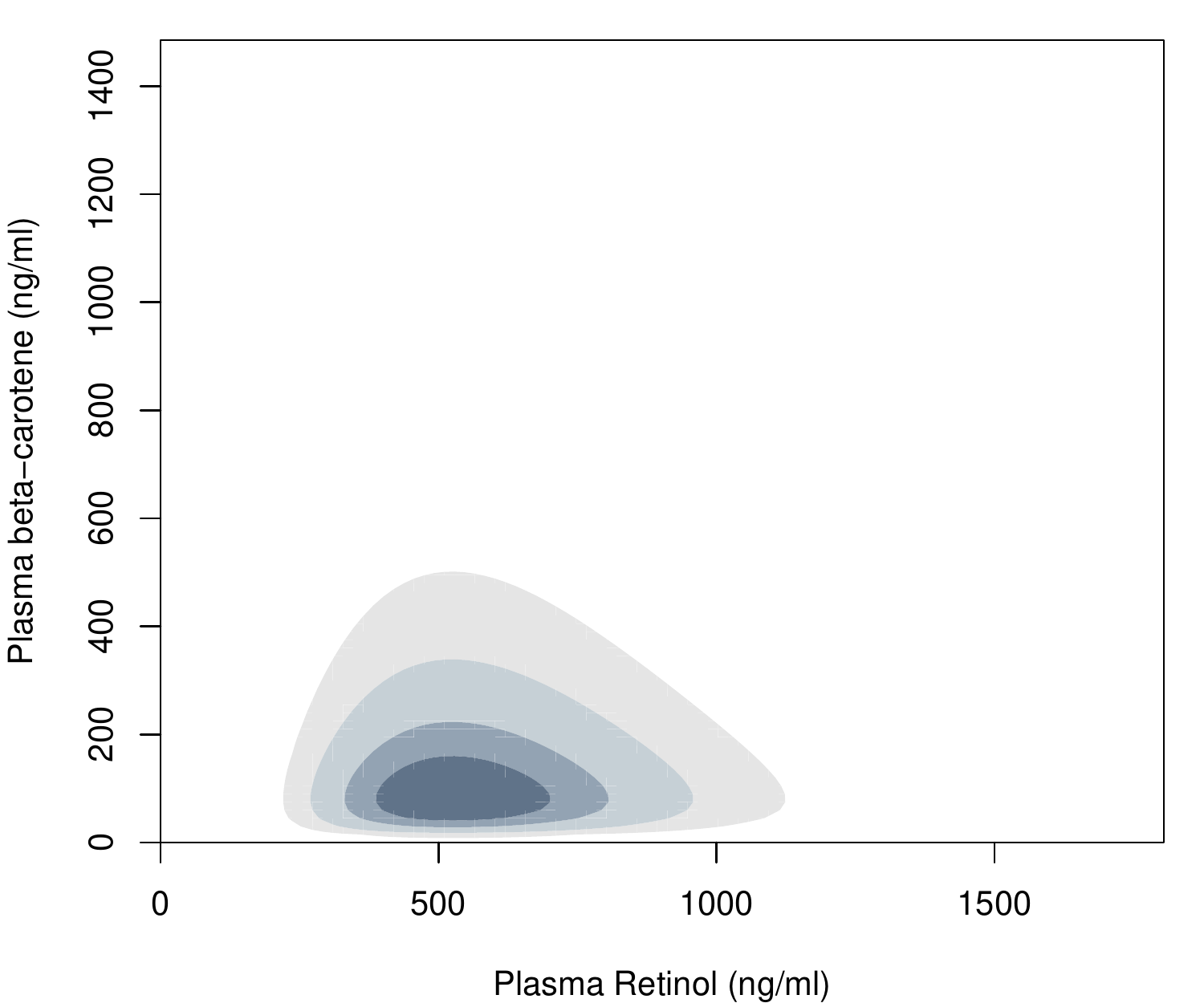}
\caption{\small Plot of data points for the plasma dataset (left panel) and the corresponding bivariate density estimated using the GMDEB approach (right panel). In the latter case, the graph shows the highest density regions corresponding to proportions $(0.25, 0.5, 0.75, 0.9)$.}
\label{fig1-2:plasma}
\end{figure}

Consider the data from a study on the association between the low plasma concentrations of retinol, beta-carotene, or other carotenoids on the increased risk of developing certain types of cancer \citep{Nierenberg:etal:1989}. The joint distribution of plasma Retinol (ng/ml) and plasma beta-carotene (ng/ml) is bounded below at zero for both variables. 
The left panel of Figure~\ref{fig1-2:plasma} shows the scatterplot of data points observed on 314 patients. 

If the bivariate density is estimated using the the standard GMM, the model with the largest BIC $= -8101.773$ has 3 components and unconstrained covariance matrix (VVV). This relatively large number of components is related to the presence of a strong skewness in the data distribution. Furthermore, a non-negligible mass of density is assigned to negative values of plasma beta-carotene.

Both issues can be solved using the proposed range-transformation approach. 
The selected model for the bivariate density estimation has BIC $= -8044.852$, with a diagonal equal variance structure and a single component (EII,1). The transformation parameters are estimated as $\hat{\lambda} = (0.155, 0.0295)$. 
The right panel of Figure~\ref{fig1-2:plasma} shows the highest density regions \citep[HDRs; ][]{Hyndman:1996} corresponding to proportions $(0.25, 0.5, 0.75, 0.9)$. In this case the joint data distribution appears to be well approximated by the estimated density. 
 
\subsection{C-horizon layer of the Kola data}

The Kola Ecogeochemistry Project (1993-1998) collected data on more than 50 chemical elements on four different primary sample materials: terrestrial moss, and the O-, B-, and C-horizon of podzolic soils located in parts of northern Finland, Norway and Russia. The main aim of the project was the documentation of the impact of the Russian nickel industry on the Arctic environment. The data are available on \citet{Reimann:etal:1998}, see also \citet{Reimann:etal:2011}. Here we analyse the distribution of nickel (Ni), copper (Cu), and chromium (Cr) on the C-horizon layer. For each of the 605 sites, the detected concentrations of the above mentioned heavy metals are provided. Clearly, concentrations are bounded below at zero, and a preliminary data exploration suggests that the joint distribution is highly skewed.

\begin{figure}[htb]
\centering
\includegraphics[width=0.8\textwidth]{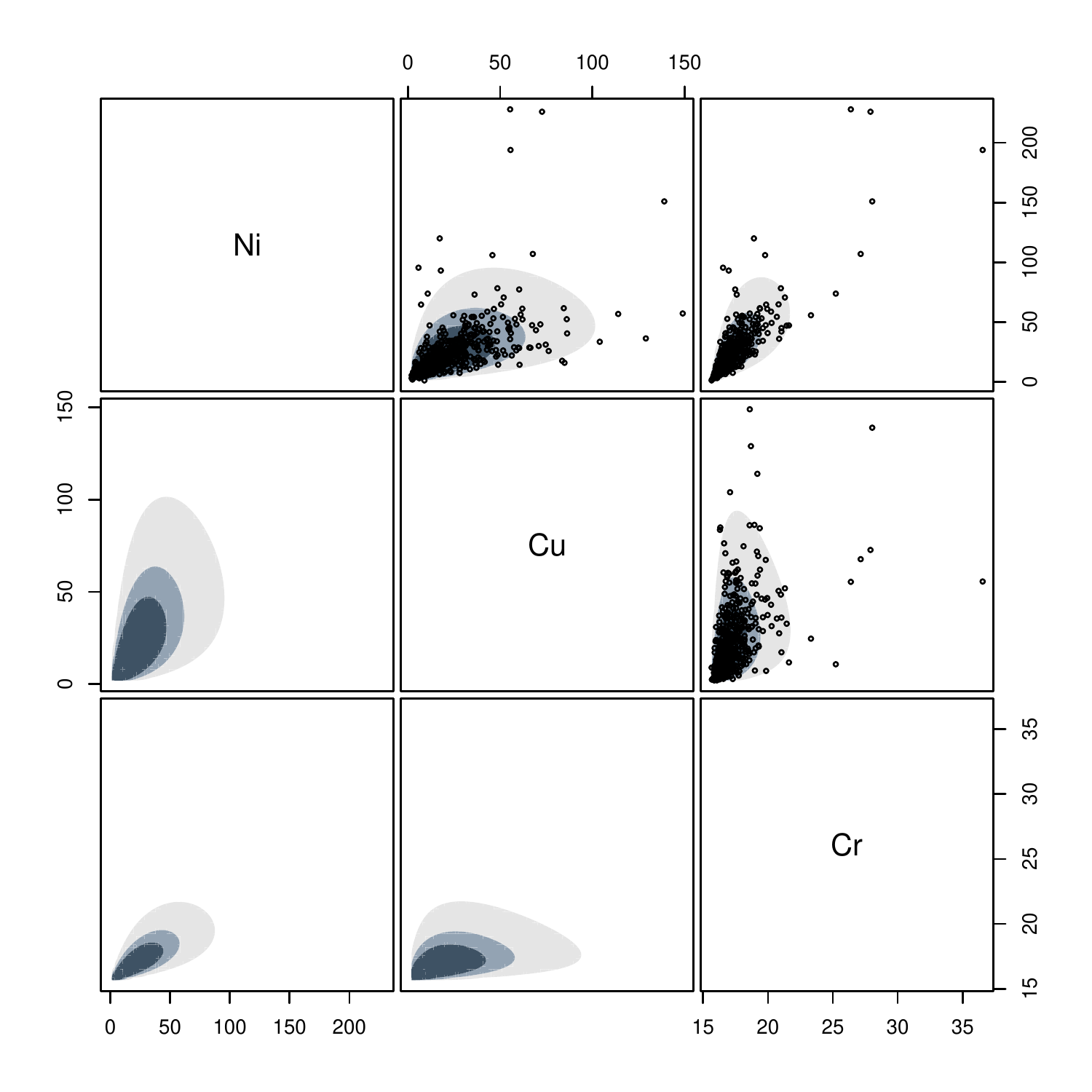}
\caption{\small Scatterplot matrix of chromium (Cr), copper (Cu), and nickel (Ni) concentrations on the C-horizon layer. The density estimated using the GMDEB approach is shown using HDRs corresponding to 25\%, 50\%, and 75\% probability regions.}
\label{fig1:chorizon}
\end{figure}

\begin{figure}[htb]
\centering
\includegraphics[width=0.7\textwidth]{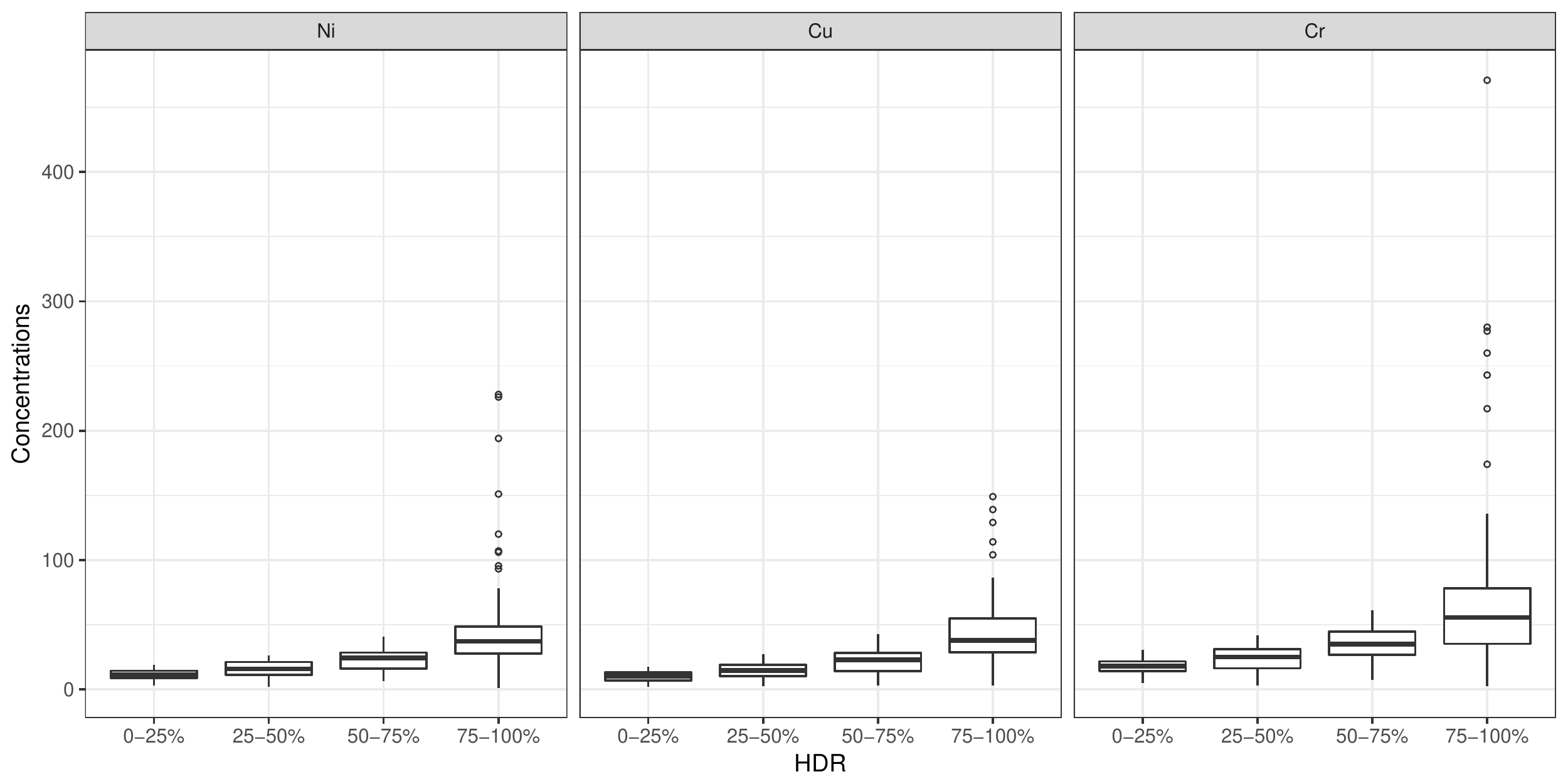}
\caption{\small Boxplots for the distributions of heavy metals conditional on HDR regions from the GMDEB density estimate.}
\label{fig2:chorizon}
\end{figure}

The selected GMDEB model according to the BIC criterion is a two components mixture model with variable volume and equal shape and orientation, i.e. VEE in \texttt{mclust} nomenclature. The estimated vector of transformation parameters is $\hat{\lambda} = (-0.0384, 0.0010, -0.0975)$. 
Figure~\ref{fig1:chorizon} contains the scatterplot matrix of heavy metal concentrations with the estimated density projected onto the marginal bivariate subspaces. The latter are shown as HDRs corresponding to 25\%, 50\%, and 75\% probability regions.
The distribution of nickel, copper, and chromium on the C-horizon layer is clearly skewed, with most sites having concentrations close to the origin. However, there are also a number of sites with relatively high concentrations. 
Further insights can be obtained by examining the distribution of metal concentrations conditional on the HDR to which the observed sites belong, as shown in Figure~\ref{fig2:chorizon}. Looking at the boxplots for the conditional distributions, we can see that the central part of the distribution, i.e. that corresponding to 0-25\% HDR, is characterised by the lowest concentration levels, whereas higher concentrations of heavy metals can be found as we move to regions of lower density.

\section{Discussion}

This paper addressed the problem of density estimation using GMMs when variables are partially or completely bounded. By introducing a range-power transformation of the data, it is possible to obtain a GMM for density estimation on the transformed data, and then to derive an accurate estimate of the density on the original scale which takes into account the natural bounds of the variables. 
The proposed model is estimated by maximum likelihood using the EM algorithm. 
We showed that this transformation-based approach is able to deal with variables having either lower bounds or both lower and upper bounds, and the results obtained are often better than those provided by other methods usually based on modified versions of kernel density estimation. 

The transformation-based approach seems to be very promising and, in principle, it could be applied to other types of non-Gaussian variables, e.g. skewed variables, and for other purposes outside density estimation, for instance in clustering.
A straightforward extension to investigate is the use of other families of transformations, such as those proposed by \citet{Manly:1976} and \citet{Yeo:Johnson:2000}.
Furthermore, although the paper deals with the problem of density estimation, the proposed methodology has implications also on model-based clustering for bounded data. 
These very important issues are deferred to future works.

\baselineskip=15pt
\bibliographystyle{chicago}
\bibliography{gmdeb_arxiv}

\end{document}